%% file: main.tex
\documentclass[letterpaper,10pt,twocolumn]{article}

\input{perks.tex}

\newcommand{\poly}{{\scshape Polypath}\xspace}
\newcommand{\mono}{{\scshape Monopath}\xspace}
\newcommand{\async}{\textsf{Presto}\xspace}
\newcommand{\jit}{JITT\xspace}
\newcommand{\jep}{JEP\xspace}
\newcommand{\pos}{PAR\xspace}
\newcommand{\pen}{pen\xspace}
\newcommand{\tp}{touch prediction\xspace}
\newcommand{\manager}{path manager\xspace}

\newcommand{\route}{I2D path\xspace}
\newcommand{\Route}{I2D Path\xspace}

\newcommand{\vsync}{sync pulse\xspace}
\newcommand{\vsyncs}{sync pulses\xspace}

\begin{document}

\date{}

\title{\vspace{-0.3in}\poly: Supporting Multiple Tradeoffs for Interaction Latency}
\author{
	\textit{Technical Report 2016-08}\\
	Min Hong Yun, Songtao He, and Lin Zhong\\
	Rice University, Houston, TX
}
\maketitle
\thispagestyle{empty}
\input{introduction}

\input{understanding}

\input{polypath}

\input{presto}

\input{implementation}

\input{evaluation}

\input{related}

\input{discussion}

\newpage
\balance
\bibliographystyle{abbrv}
\bibliography{bib/ref}
\end{document}

%% file: perks.tex
\usepackage{times,epsf,epsfig,textcomp,amsmath,amssymb,wrapfig,moreverb}
\usepackage{comment}
\usepackage{array}
\usepackage{colortbl}
\usepackage{url}
\usepackage{color,soul}
\usepackage[dvipsnames]{xcolor}
\usepackage{graphics}
\usepackage{multirow}
\usepackage{capt-of}
\usepackage{siunitx}
\usepackage{listings}
\usepackage{courier}
\usepackage[normalem]{ulem}
\usepackage{setspace}
\usepackage{caption}
\usepackage{placeins}
\usepackage{afterpage}
\usepackage[boxed]{algorithm2e}
\usepackage{xspace}
\usepackage{hhline}
\usepackage{hyperref}
\usepackage{leading}
\usepackage{balance}
\usepackage{geometry}
\usepackage{lipsum}
\usepackage{titlesec}
\usepackage{enumitem}
\usepackage{graphicx}
\usepackage{subfig}
\usepackage{subfloat}

\titlespacing\section{0pt}{8pt plus 4pt minus 2pt}{4pt plus 2pt minus 2pt}
\titlespacing\subsection{0pt}{6pt plus 4pt minus 2pt}{2pt plus 2pt minus 2pt}
\titlespacing\subsubsection{0pt}{4pt plus 2pt minus 2pt}{2pt plus 2pt minus 2pt}

\usepackage{tikz}

\definecolor{lightgray}{gray}{0.9}
\definecolor{lightblue}{rgb}{0.9,0.9,1}
\definecolor{red}{rgb}{1,0,0}
\definecolor{darkgreen}{rgb}{0.4,0.7,0.3}

\newcommand{\state}[1]{\textsc{#1}}
\newcommand{\program}[1]{\textsf{\small #1}}
\newcommand{\code}[1]{\texttt{\small #1}}

%% file: introduction.tex
\subsection*{Abstract}
Modern mobile systems use a single input-to-display path
to serve all applications. In meeting the visual goals
of all applications, the path has a latency inadequate 
for many important interactions.
To accommodate the different latency requirements and visual constraints
by different interactions, we present \poly, a system design in which application
developers (and users) can choose from multiple path designs for their application
at any time. 
Because a \poly system asks for two or more path designs, we 
present a novel fast path design, called \async. 
\async reduces latency by judiciously allowing frame drops and tearing.

We report an Android 5-based prototype of \poly with two path designs: Android legacy and \async.
Using this prototype, we quantify the effectiveness, overhead, and user experience of \poly, especially \async,
 through both objective measurements and subjective user assessment. 
 We show that \async reduces the latency of legacy touchscreen drawing applications by almost half; 
 and more importantly, this reduction is orthogonal to that of other popular approaches and is
 achieved without any user-noticeable negative visual effect. 
 When combined with \tp, \async is able to reduce the touch latency below \SI{10}{ms}, 
 a remarkable achievement without any hardware support. 

\section{Introduction}\label{sec:intro}
\begin{figure}
	\centering
	\subfloat[Stock Android \label{sfig:testa}]{
		\includegraphics[width=.4\linewidth]{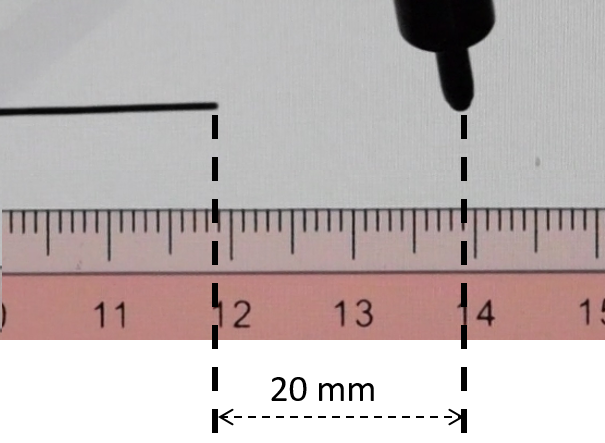}
	}\hspace{0.5cm}
	\subfloat[\async \label{sfig:testb}]{
		\includegraphics[width=.4\linewidth]{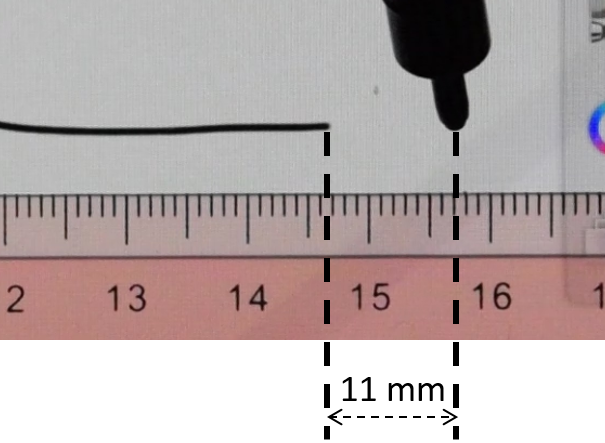}
	}
	\caption{Touchscreen drawing translates latency, e.g., \SI{82}{ms}, into a visible gap, e.g., \SI{20}{mm}, as the pen moves and the line head falls behind (\program{Autodesk Sketch}, camera captured). }
	\label{fig:touchdelay}
\end{figure}

The input-to-display path, or \emph{\route} for short, is an important operating system (OS) service because it 
determines the user-perceived latency of interaction.
Today's mobile OSes employ the same path design for all interactions, a system design we call \emph{\mono}.
This path design therefore must meet the visual goals of all applications: consistent frame rate,
no frame drops, and no tearing effects. It achieves this with a coarse-grained design at the cost of long latency,
over \SI{60}{ms}~\cite{agawi,lesnumeriques_21devices,lesnumeriques_note3}. 
Although this latency may be fine for point/selection-based interactions~\cite{seow2008designing},
it is annoying for others.
For example, touchscreen-based drawing and dragging manifest latency as a spatial gap between the touch point and the visual effect~\cite{1msdelay}; a latency of \SI{60}{ms} produces an obvious, annoying gap, as illustrated in Figure~\ref{fig:touchdelay}.
The same is true for augmented-reality interactions based on head-mounted displays. The original motivation of our work is 
to reduce the latency for such demanding interactions.

We quickly realized that latency reduction is not free. Without specialized, 
expensive hardware like that used in~\cite{ng2012designing}, one has to make tradeoffs between the visual goals and latency. 
Luckily, we find these goals are not necessary or can be relaxed for many interactions under
modern hardware and software. 
All these point to the \mono design of modern mobile OSes as
the fundamental problem. To cater to the different latency requirements and visual constraints of diverse interactions,
we argue that mobile OSes should follow \emph{\poly} and support multiple \route designs in the same system.

Our guiding principles for \poly are twofold: (\textit{i}) different applications and different parts of an application can employ different path designs; and (\textit{ii}) application developers and users should decide which part of an application employs which path design.

In this paper, we report our design of \poly that supports unmodified legacy applications. 
The key insight behind the design is that the interface between an application and the rest of the \route is clean and independent; each group of events delivered by the input to the application can be handled by a given path design, independently of the group before or after. Our \poly system provides an asynchronous API for developers and users to bind a path design to an application; it further ensures that an input event only experiences one path design, a property we call \emph{path integrity}.

Because a \poly system asks for two or more path designs, we provide a novel \route design, called \async. 
Compared to the path used in today's \mono systems, \async almost halves the latency by judiciously allowing frame drops and tearing, making a very different tradeoff.
In particular, \async overcomes the coarse granularities of the legacy path design through two key techniques. \emph{Just-in-time trigger} eliminates strict synchronization in the path with the display.
\emph{Just enough pixels} allows the \route to operate on only updated pixels, or dirty regions, of a frame.

We report an Android 5-based implementation of \poly that supports two path designs: Android legacy and \async. We evaluate its effectiveness in latency reduction, overhead, and user experience with both objective measurements and subjective assessment. Our measurements show that \async reduces the latency by \SI{32}{ms} on average for top drawing applications from Android Play Store, with a power overhead that can be eliminated with SDK support. 
The effectiveness is obvious from Figure~\ref{fig:touchdelay}. Importantly, we show that the latency reduction resulting from \async is orthogonal to that from known techniques such as \tp used by iOS 9. 
When combined with \tp, \async is able to reduce the touch latency below \SI{10}{ms}, 
 a remarkable achievement without any hardware support. 
Double-blind user evaluation demonstrates that for the drawing 
applications tested, \async improves the user experience without noticeable side effects.

\vspace{+0.5ex} In summary, we make the following contributions:

\begin{itemize}[leftmargin=*]

\vspace{-1ex}\item We present a general model for the \route and identify that coarse granularity in today's \route design contributes significantly to the interaction latency. We show that this legacy design sacrifices latency to strictly meet several visual goals that are not always necessary today.

\vspace{-0.5ex}\item We provide a design and implementation of \poly that supports multiple \route designs in the same system. \poly allows application developers and users to make different tradeoffs for latency without affecting other applications.

\vspace{-0.5ex}\item As part of our \poly system, we present \async, a novel \route design that reduces latency by almost half. We provide a prototype implementation of \async based on Android 5 that supports unmodified legacy applications and evaluate it with both objective measurements and subjective assessment. 
\end{itemize}

%% file: understanding.tex
\section{\Route and Its Tradeoffs}\label{sec:understanding}
In this section, we present a conceptual model for the \route to understand its design tradeoffs. 
We show that the \route design of today's \mono systems represents a specific tradeoff point in a large design space between latency and other computing goals.
By showing many other possible, desirable tradeoff points, we motivate the need for \poly operating systems in which multiple \route designs are supported.

\subsection{\Route Model}\label{sec:path}

\begin{figure}[t]
		\centering
		\includegraphics[width=1\linewidth,trim={3cm 3cm 7cm 2cm},clip]{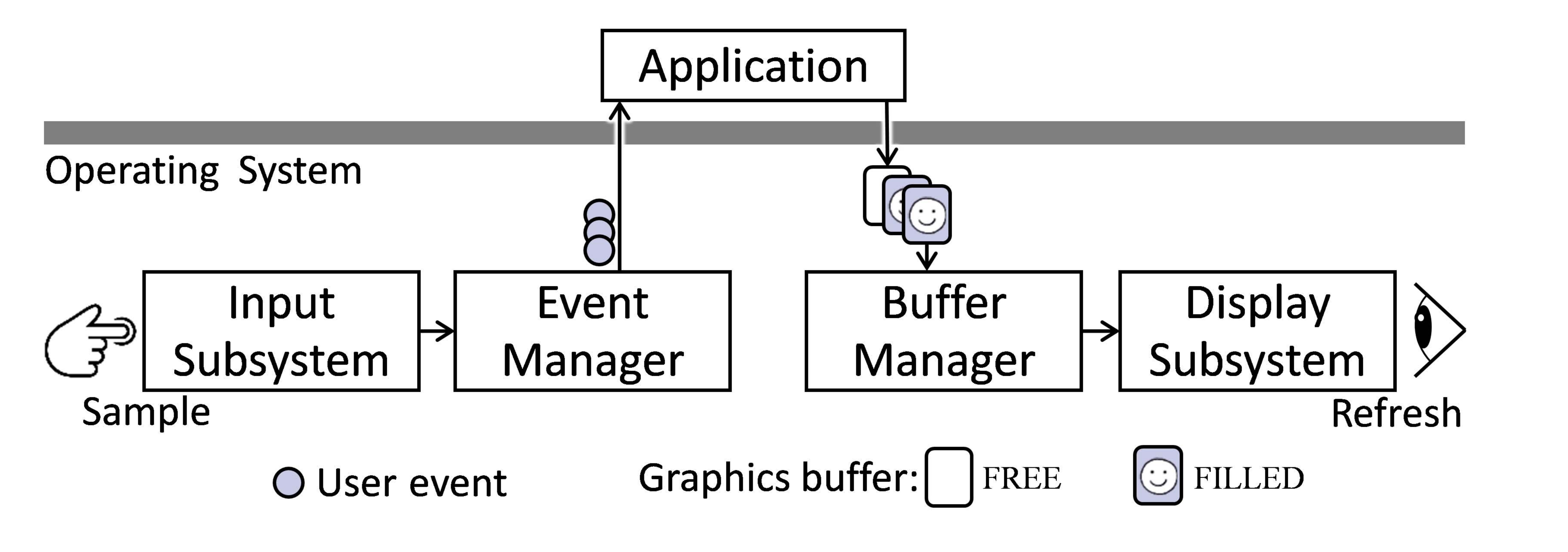}
		\caption{ 
			A general model for the \route: the event manager batches events from the input subsystem and delivers them to the application; the application then draws a frame on a buffer supplied by the buffer manager.
			The buffer manager transfers buffer ownership to the display subsystem. 
			}
		\label{fig:io_path}
\end{figure}

Based on an understanding of mainstream mobile OSes, i.e., iOS and Android, we devise a five-part model for the \route as shown in 
Figure~\ref{fig:io_path}. The model includes input, event manager, application, buffer manager, and display. All except application are part of the OS (not necessarily in kernel space though). 

The \emph{input} subsystem includes the input device driver. 
It samples the physical world and produces software events.
The sampling rate is typically \SI{120}{Hz}~\cite{iosPrediction} but can range from \SI{80}{} to \SI{240}{Hz}.

The \emph{event manager} is per-application. It buffers events from the input and delivers them to the application. The buffering is necessary because the input subsystem produces the events faster than the display refreshes. High-rate events are necessary because of application's desire for smooth visual effects.

The \emph{buffer manager} is also per-application. It manages the application's \emph{graphics buffers}. 
The application processes the input events, takes a \state{free} buffer, marks it \state{busy$_{app}$}, draws a frame on it and then marks it \state{filled}.

The \emph{display} subsystem includes the software part of the composer.
It takes \state{filled} buffers from multiple applications, marks them as \state{busy$_{disp}$}, and handles them
to the hardware, which composes the buffers and sends the composition to the display panel serially. After that, the display subsystem marks the buffers as \state{free}.
Because composing is done by specialized hardware, it adds  negligible latency.
The display controller refreshes the display panel and fires a \vsync periodically, with the period of {\small $T_{sync}$}. 
In modern mobile systems,  {\small $T_{sync}$} is typically \SI{1/60}{s}~\cite{graphicxArch, iosPrediction}.

\input{latency}

\input{tradeoff}

%% file: latency.tex
\subsection{Coarse-Grained I2D Path Design}\label{sec:monopath_latency}

\begin{figure}[t]
		\vspace{0pt}
		\centering
		\includegraphics[width=0.9\linewidth,trim={5cm 0cm 2cm 0}]{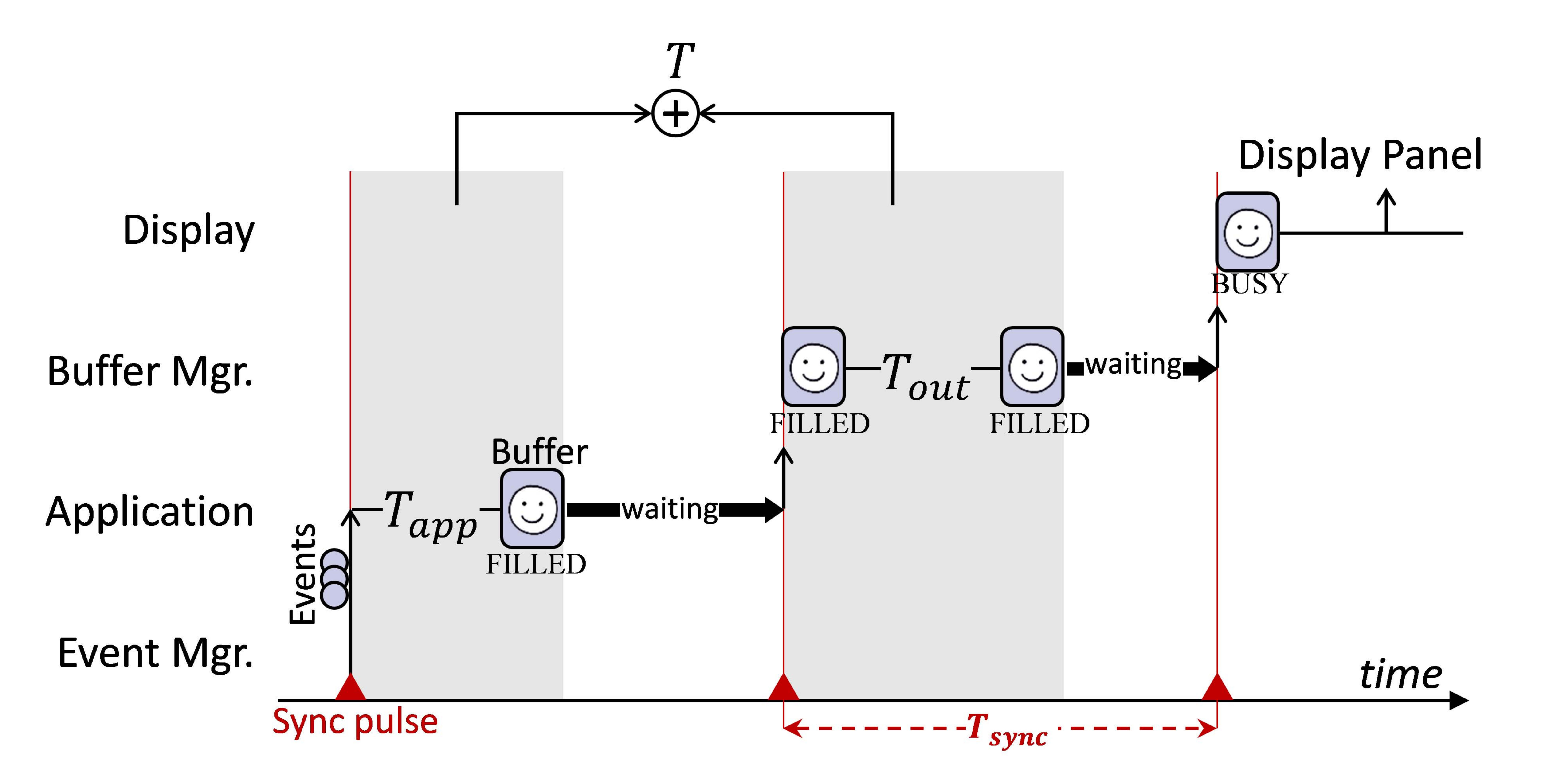}
		\caption{The timeline for the legacy \route in action: the \vsync fired by the display controller triggers the event and buffer managers, causing waiting and delay.
			}
		\label{fig:latency_path_origin}
\end{figure}

At the cost of long latency, the \route in today's mobile OSes guarantees three visual goals: a consistent frame rate, no frame drops, and no tearing effects. 
It achieves it with a design that is coarse-grained in both time and space. In time, its event and buffer managers strictly synchronize 
with the sync pulse produced by the display controller; in space, it assumes a buffer can not be read and written at the same time. 
Figure~\ref{fig:latency_path_origin} provides a timeline for the legacy \route. 

\textbf{Synchronization}:~~
In the legacy design, both the event and buffer managers synchronize with a periodical signal, a.k.a. the \vsync, fired by the display subsystem when refreshing. 
The event manager waits for a \vsync to deliver buffered events to the application.
Assume the application takes {\small $T_{app}$} to process the events, produce a frame and write it into a graphics buffer. It will wait another {\small $(T_{sync}-T_{app})$} until the next \vsync so that the buffer manager can process the buffer. 
Therefore, synchronization at the event manager introduces an average latency of {\small $(T_{sync}-T_{app})$}. Android reduces this by triggering the event manager \SI{7.5}{ms} after the \vsync~\cite{graphicxImpl}. In this case, the latency would be {\small $(T_{sync}-T_{app}-7.5_{ms})$}.

The buffer manager waits for a \vsync to change graphics buffers' ownership among the application, display subsystem and itself.
Assuming this process takes {\small $T_{out}$}, this synchronization introduces an average latency of {\small $(T_{sync}-T_{out})$} because the buffers will be externalized only at the next \vsync. 
These synchronizations together ensure a \emph{consistent frame rate}. 
Synchronization of the buffer manager additionally ensures \emph{no frame drops}.  
No matter how quickly an application finishes drawing, the buffer manager transfers buffer ownership only on a \vsync.

\textbf{Atomic Buffer}:~~
Noticeably, the buffer manager does not give a \state{busy$_{disp}$} buffer to the application, avoiding the same buffer being read by the display and written by an application at the same time. This is sufficient but not necessary to \emph{avoid tearing effects}. 
However, the buffer manager does not have better strategy because it has no idea about which pixels have been changed from one frame to the next, i.e., dirty region. 
This strategy makes an average latency of {\small $0.5\cdot T_{sync}$} due to the display refreshing necessary because the application has to finish writing in a buffer before the display starts to externalize it.
As a result, any \state{busy$_{disp}$} buffer has to wait for the next display refreshing to be sequentially externalized, introducing an average latency of  {\small $0.5\cdot T_{sync}$}.

\vspace{+2ex}All together, we estimate the average latency due to the coarse-granularities as 
\vspace{-2mm}
\begin{equation}
\small 2.5\cdot T_{sync}-(T_{app}+T_{out})
\label{eq:latency}
\end{equation}

\vspace{-2mm}
For a typical Android application, this latency is about \SI{26.6}{ms} with {\small $T_{sync}=$ \SI{1/60}{s}} and the \SI{7.5}{ms} optimization deducted. This accounts for close to half of the latency we observe on Android devices.
One na\"ive way to reduce this latency is to simply reduce {\small $T_{sync}$}. 
However, this would incur proportionally higher power consumption by the display subsystem.
Even worse, when reducing {\small $T_{sync}$}, the subsystems will have to process proportionally faster because their processing time must be masked by  {\small $T_{sync}$}, leading to system-wide proportionally higher power consumption and the use of expensive hardware.

%% file: tradeoff.tex
\subsection{Design Tradeoffs in \Route Design}\label{sec:newrequirements}

\begin{figure}[t]
	\centering
		\includegraphics[width=1\linewidth,trim={6cm 25cm -3cm 15cm},clip]{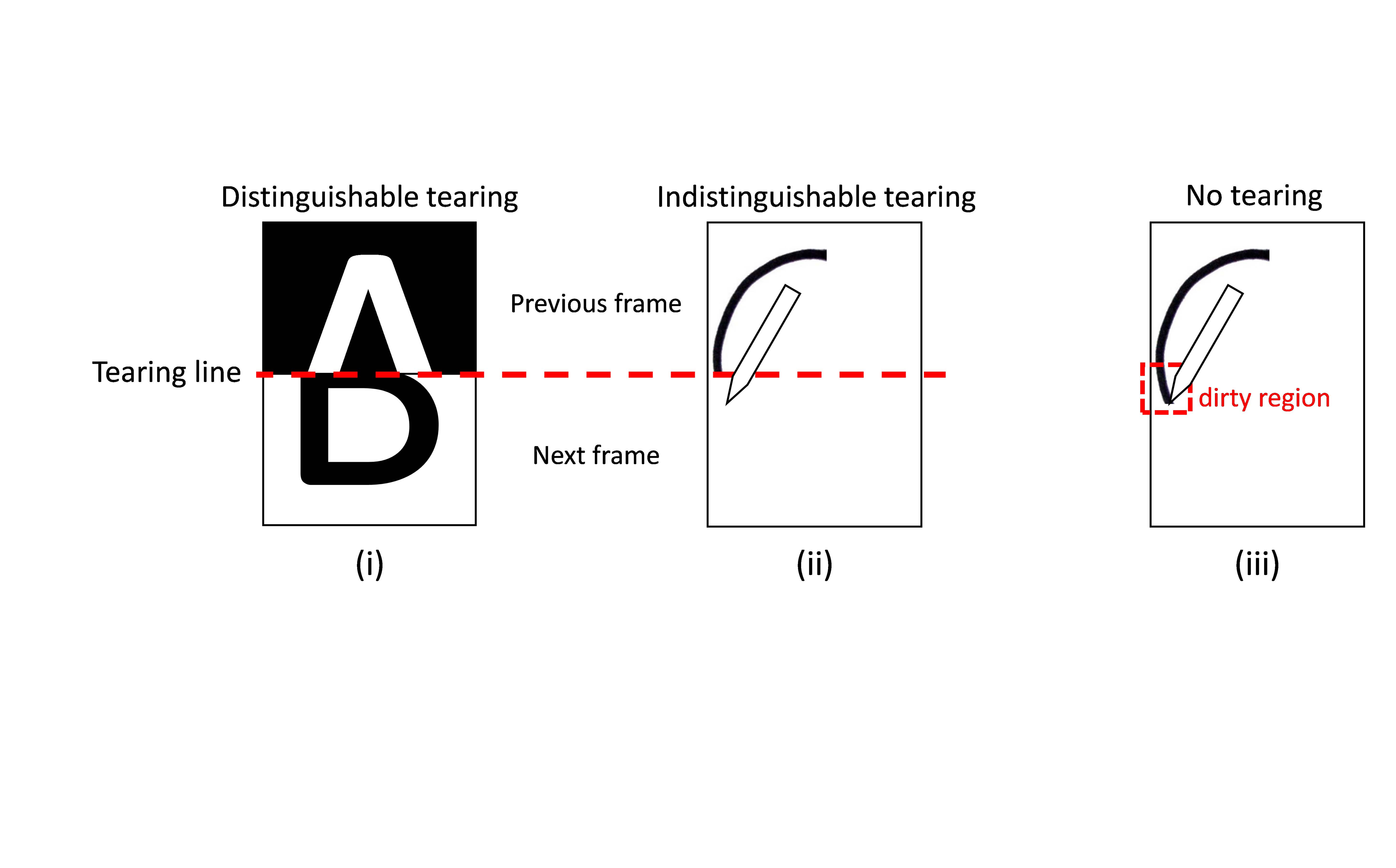}
		\caption{Tearing effect may happen when the display switches from one frame to the next in the middle of externalizing the first. As a result, the screen will show the early part of the first frame and the late part of the second, joined at the tearing line. 
			If the tearing line cuts across a large dirty region, the tearing effect can be visible and annoying as is in (i). If the dirty region is very small, like in the cases of drawing, the tearing effect is indistinguishable from the effect from latency, as is in (ii), when compared to the perfect case in (iii).}
		\label{fig:tearing}
\end{figure}

The legacy path design analyzed above represents one particular point of tradeoff between latency and other computing goals. 
In some sense, it represents an extreme point: it efficiently and strictly meets all three visual goals, at the cost of latency. For many important latency-sensitive interactions, these goals can be relaxed, especially in view of the recent hardware and software development. As a result, \emph{different tradeoffs are possible},
especially in favor of short latency.

First, HCI research has shown that different interactions have different latency requirements~\cite{Claypool:2006:LPA:1167838.1167860,Claypool:2010:LKP:1730836.1730863}. It is known that the human-perceptible threshold for causality is about \SI{100}{ms}~\cite{miller1968latency,seow2008designing}. That is, if we click a button on screen and the button changes within \SI{100}{ms}, we would barely notice the delay. As a result, the \SI{100}{ms} latency has been considered as adequate for keyboard/mouse-based interactions as well as touchscreen-based point/selection. In contrast,
Ng. et al~\cite{ng2014blink,ng2012designing} showed that the just-noticeable difference (JND) latency 
for object dragging on the touchscreen is
\SI{2}-\SI{11}{ms}, much shorter than what modern mobile systems can deliver.
Microsoft went further to argue for \SI{1}{ms} latency for touchscreen interactions~\cite{1msdelay}.
Latency reduction, however, is not free. Microsoft achieves it  \SI{1}{ms} latency
only with expensive, specialized hardware. Given the hardware, latency reduction often requires more computation, 
e.g, event prediction~\cite{lavalle2014head} and speculation~\cite{lee2015outatime}, or relaxing the visual goals, as explained below.

Second, the three visual goals met by the legacy \route are not absolute.
For some applications and interactions, they are not necessary at all, especially on modern mobile hardware and software. 
Hardware improvements, i.e., faster CPU, GPU and larger memory, have enabled a consistent frame rate of \SI{60}{fps} on modern mobile systems. 
Recent studies~\cite{claypool2007frame,claypool2006effects} have shown that users cannot perceive changes in frame rate when it is above \SI{30}{fps}. 
Similarly, frame drops can be allowed if they are not consecutive and the frame rate is kept above \SI{30}{fps}.
For another example, drawing on touchscreen usually has visual effects limited to the touched position. 
Tearing effects would be barely noticeable by human eyes or even high-speed cameras. 
Indeed, they are almost indistinguishable from the effect of latency as highlighted by Figure~\ref{fig:tearing}.

Moreover, it can be profitable for user experience to trade these visual goals for shorter latency. 
Janzen and Teather~\cite{janzen201460} showed that latency affects user performance with 
touchscreen interaction more than frame rate does. Our fast path design, \async, to be presented in \S\ref{sec:rapido},
also carefully drops delayed frames in order to cut overall latency. 
We believe that it should be up to the application developers and users to determine what tradeoffs are
profitable. 

Finally, on battery-powered mobile devices, the visual goals may be traded for lower power consumption. For example, on Nexus 6, lowering the frame rate from \SI{60} to \SI{30}{fps} reduces the overall system power consumption by \SI{300}{mW}, or 20\%, when running \program{Angrybird}.

\vspace{+1ex}In summary, hardware and software advancements described above make more tradeoffs 
between latency and other computing goals possible.  
Because the latency of today's \mono system is inadequate for many interactions, we argue 
for a \poly system design in which multiple path designs making different tradeoffs for latency coexist. 
In a \poly system, application developers and users can decide when to apply which path design to 
which application.

%% file: polypath.tex
\section{Design of \poly}\label{sec:poly}

Because all existing mobile OSes are \mono, we face many important decisions when designing \poly.
 In this section, we elaborate these decisions.

By introducing multiple path designs, \poly first faces the problem of naming and binding. That is, in a \poly system, an application must be able to \emph{name} a path design and be able to use it by \emph{binding} to it. Indeed, much of our \poly design involves answering questions about the naming and binding. In this section, we speak of binding in an abstract way because different systems may implement it differently.

First, we must support legacy applications that are designed with the \mono system in mind. 
This is possible via two design choices. (\textit{i}) First, as apparent from Figure~\ref{fig:io_path}, 
there are two interfaces between an application and the rest of path: event delivery from the 
event manager and the buffer exchange with the buffer manager. 
As long as these two interfaces are kept unchanged, a path design should support unmodified 
legacy applications. (\textit{ii}) Moreover, the naming 
of path and the binding between path and application must be achieved outside the application. 
In our \poly design, they are realized by an OS module, called \emph{\manager}.
The \manager records the path preference of each application during app installation 
and updates the record via a system API it exports. 
The API has the simple form of \code{ApplyPath(app\_name, path\_name)}.
Importantly, the decision to support legacy applications further allows us to focus on the OS part of \route,
which includes the event and buffer managers. As a result, when we say \emph{path design} in this work, 
we are referring to the design of the path, event and buffer managers.

Second, we must decide on the lifetime of the binding between an application and a path design. 
That is, when is a binding created and when does it change? In one extreme, the path manager can decide binding for each sequence of events delivered to the application and as a result, can change the binding from one sequence of events to the next. 
We consider this fine granularity as unnecessary. 
Instead, we opt for bind-by-need strategy similar to call-by-need evaluation~\cite{henderson1976lazy}. 
That is, once a binding is created, at the time of the application launch, it lasts until the application or the system explicitly asks for a change, via the path changing API. This design invokes the path manager much less frequent and therefore has higher efficiency and better reliability.

Finally, when and how can the path binding of an application be changed? We note that the invocation of \code{ApplyPath()} can be asynchronous, i.e., it can happen any time during the application's lifetime. 
Notably, the path design consists of two disjoint parts, linked by the application: the \emph{event manager} and the \emph{buffer manager}. If the path design is changed after the event manager delivers the events to the application but before the buffer manager gives the application a buffer, the events will experience the event and buffer managers from two different designs. This behaviors violates the expectation of developers and users that an event should experience the same path design, a property we call \emph{path integrity}. 
Therefore, at the invocation of \code{ApplyPath()}, if the event manager has already buffered events, the \manager ensures that these events follows the current path; otherwise, it binds the new path to the application for the incoming events.

We note that this \poly system design naturally supports any number of paths to be added or removed from the system. When the system cannot find the path required by an application, it can fall back to a default. 
Additionally, applications using the same path design in \poly are as isolated from each other as they would be in the \mono system.

%% file: presto.tex
\section{\async: A Fast Path Design}\label{sec:rapido}
Because \poly asks for two or more \route designs and existing mobile OSes only provides one, we next present a novel \route design, called \async.
Compared to the path used in today's \mono system, \async almost halves the latency by making a different tradeoff between latency and other computing goals. In particular, \async eliminates the coarse granularities of the legacy path design, as discussed in \S\ref{sec:monopath_latency}, through two key techniques, \emph{just-in-time trigger}, or \jit, and
\emph{just-enough pixels}, or \jep. Speaking of tradeoff, \async judiciously allows frame drops (\jit) and tearing (\jep) in favor of short
latency. 

 \jit eliminates the synchronization of the event and buffer managers.
It aims to get as many input events to the application as the resulting frame will be ready by the next display refresh. 
The \jit buffer manager transfers the buffer ownership to the display subsystem immediately after the application finishes drawing, without waiting for a \vsync. 

\jep and its approximation, \emph{position-aware rendering}, or \pos, further alleviate the atomic use of buffers by judiciously allowing an application to write into a \state{busy$_{disp}$} buffer that is being externalized by the display.

\input{jit}

\input{jep}

%% file: jit.tex
\subsection{\jit: Just-In-Time Trigger}\label{sec:design_jitt}
\jit removes synchronization in the event and buffer managers.
With \jit, the event manager judiciously decides when to deliver buffered events to the application; and the buffer manager transfers buffer ownership as soon as the application finishes drawing, without waiting for the \vsync. 
Ideally, the buffer manager would deliver the buffer filled by the application's response right before the next display refresh.
Recall that we denote the time it takes the application to process the events and fill the buffer as {\small $T_{app}$}, the time it takes the buffer manager to transfer the buffer ownership to the display subsystem as {\small$T_{out}$}. 
For brevity, we denote {\small $(T_{app}+T_{out})$} as {\small$T$}.

\begin{figure}[t]
		\centering
		\includegraphics[width=1\columnwidth]{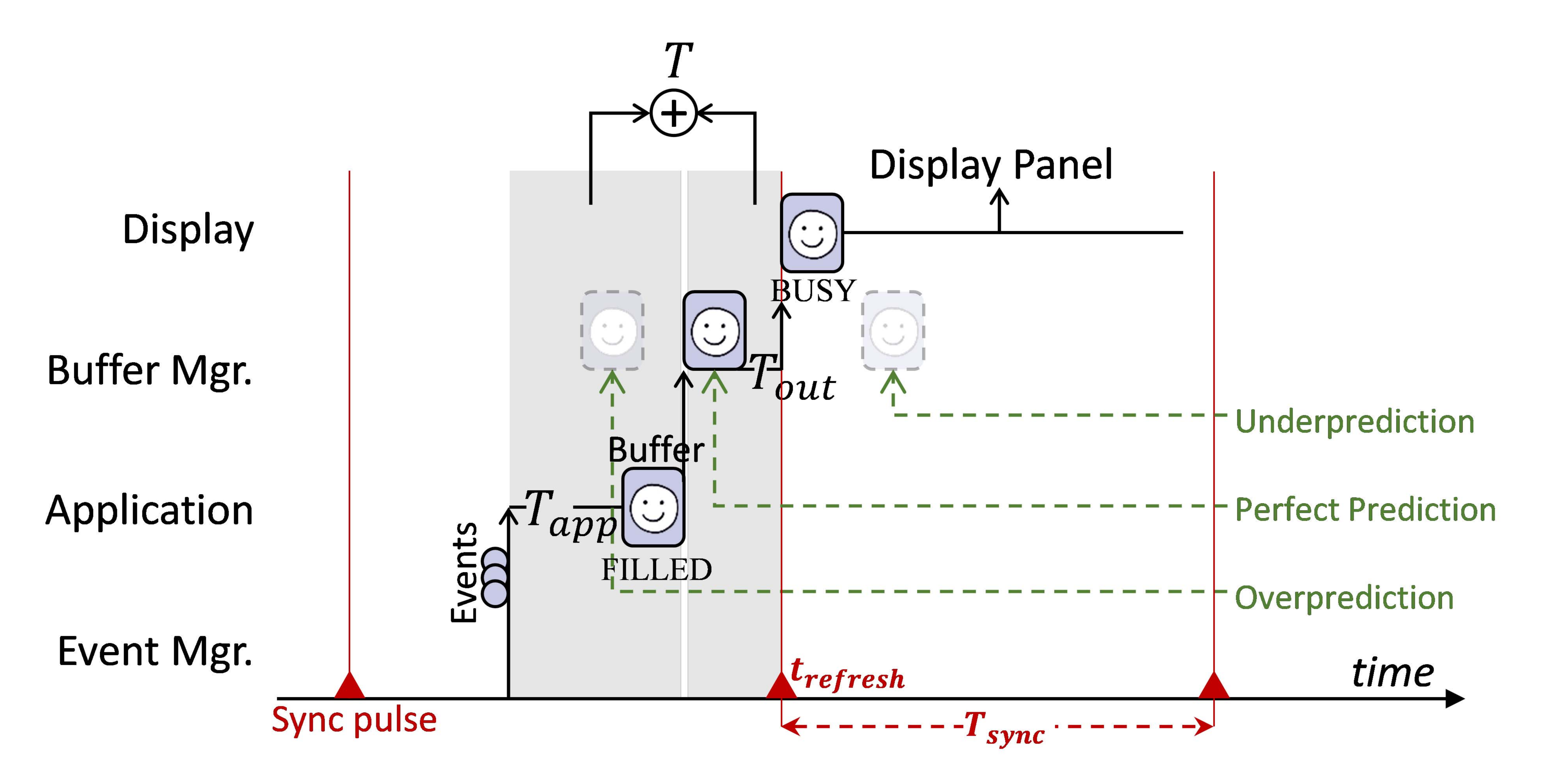}
		\caption{\jit removes synchronization in the event and buffer managers. It decides when the event manager delivers events to the application so that the buffer manager would deliver the buffer filled by the application right before the next display refresh. To do so, it must predict how long it will take from the event delivery to the buffer delivery, or {\small$T$}. Overprediction {\small$(T'>T)$} leads to an increase in latency by {\small$(T'-T)$}; underprediction {\small$(T'<T)$} leads to an increase in latency by {\small$T_{sync}$}.}
		\label{fig:jitt}
\end{figure}

In the ideal case with \jit, no events would have to wait more than {\small$(T+T_{sync})$} for their application response to externalize, with average being {\small$(T+0.5\cdot T_{sync})$}. 
This is illustrated by the perfect prediction path in Figure~\ref{fig:jitt}.
Therefore, knowing when the display refreshes next, denoted by {\small$t_{refresh}$}, \jit must predict {\small$T$}, and let the event manager deliver the events at {\small$(t_{refresh}-T')$} where $'$ indicates prediction.

{\small$T_{app}$} and {\small$T_{out}$} can be easily predicted using history.
Much of the prediction algorithm is system-specific and we will revisit when reporting the implementation (\S\ref{sec:impl_presto_jitt}).
Below, we focus on one important design issue.
Inaccurate prediction increases latency of \jit.
An overprediction {\small$(T'>T)$} makes the event manager deliver events too soon. 
That is, if events arrive between {\small$(t_{refresh}-T')$} and {\small$(t_{refresh}-T)$}, the corresponding frame would wait for the screen refresh and increase the average latency by {\small$(T'-T)$}.
This is illustrated by the overprediction path in Figure~\ref{fig:jitt}.
An underprediction {\small$(T'<T)$} makes the event manager wait too long to deliver the buffered events and as a result, the buffer manager will not be able to transfer the resulting graphics buffer to the display subsystem by the next display refresh, adding an entire {\small$T_{sync}$} to the average latency. 
This is illustrated by the underprediction path in Figure~\ref{fig:jitt}.
Apparently, the latency penalty is significantly higher in the case of underprediction. 

\jit copes with underprediction in two ways. First, it favors overprediction between overprediction and underprediction. That is, it looks for the upper end when using history. Moreover, with prediction {\small$T'$}, instead of triggering the event manager at {\small$(t_{refresh}-T')$},  \jit calculates when the last event would arrive before {\small$(t_{refresh}-T')$} and triggers the event manager when this event arrives. This trick essentially 
adds a variable offset to {\small$T'$} in favor of overprediction. 
Second, \jit recovers from underprediction by dropping the frame in the buffer delayed due to underprediction. Importantly, this recovery mechanism does not drop two frames in a row. When underprediction happens, the buffer manager will have two \state{filled} buffers when \jit triggers it:  one delayed and the other newly produced.
Then, the buffer manager drops the older buffer by marking it as \state{free} and transfers the newer one to the display subsystem.
If \jit underpredicts one more {\small $T$} in a row, the buffer manager does not drop the delayed frame anymore but propagates the delay until no underprediction happens or the application stops producing frames.
The worst case is when {\small$T_{app}$} changes abruptly and the \jit buffer manager drops every other frame, the frame rate becomes half, or \SI{30}{fps} on modern mobile systems.

%% file: jep.tex
\subsection{\jep: Just-Enough Pixels}\label{sec:design_jep}

As explained in \S\ref{sec:path}, in modern mobile systems, when an application requests a graphics buffer, the buffer manager will give it a \state{free} one. Therefore, the application cannot write into the \state{busy$_{disp}$} buffer that is being externalized by the display subsystem. This atomic buffer access avoids tearing but adds a latency of $0.5\cdot T_{sync}$ on average as discussed in \S\ref{sec:monopath_latency}. \jep reduces this latency by judiciously allowing the application to write into the \state{busy$_{disp}$} buffer, without tearing.

\jep leverages partial-drawing APIs like~\cite{segal1992opengl, glsurfaceview} and a modern mobile display trend~\cite{191579, dsiCommand}: an in-display memory from which the display panel reads pixels, not directly receiving from the composer. 
The key idea is to make the atomic area smaller, i.e., the dirty region of the new frame, and let the display subsystem take only the dirty region to compose and update
the in-display memory only before the display panel starts externalizing the dirty region.
This is possible without tearing because a modern display externalizes a frame sequentially, pixel by pixel and updating only the dirty region reduces the memory copy between the buffer and the display subsystem, e.g., by \SI{7178.0}{KB/s}~\cite{191579}.

Specifically, \jep needs to answer two questions: (1) where is the starting point of the dirty region? That is, in how many pixels will the display externalize before reaching the dirty region? (2) how fast is the display subsystem externalizing pixels?
The use of partial-drawing APIs answers (1).
The answer to (2) is independent of applications and can be accurately profiled. For example, in our prototype, we find the display subsystem externalizes \SI{221}{M} pixels per second.

Because most legacy mobile applications do not use the partial-drawing APIs and not all mobile displays feature the internal memory,
we next present \pos, an approximation of \jep, to support legacy applications and displays.

\input{position}

%% file: position.tex
\subsubsection{\pos: Position-Aware Rendering}\label{sec:design_position}

To support legacy applications and displays, \pos allows an application to write in the \state{busy$_{disp}$} buffer that is being externalized by the display subsystem.
To minimize the risk of tearing effects caused by concurrent buffer accesses, 
\pos must be confident that the application would finish writing into the buffer BEFORE the display subsystem starts externalizing a dirty region. 
Therefore, in addition to the previous two questions to \jep, \pos must answer a third question: how long will it take the application to finish drawing into the buffer?
Notably, the answer is essentially {\small $T_{app}$} of which a prediction is available from \jit as described in~\S\ref{sec:design_jitt}. Like in \jit, underprediction is more harmful than overprediction in \pos: underprediction risks tearing effects while overprediction only decreases latency reduction.

To further limit tearing effects, we exploit the fact that many applications will have visual effects and henceforth dirty regions limited to around the touched position; and tearing in this area is barely distinguishable from effect of latency as shown in \S\ref{sec:newrequirements}. \async will apply \pos only if the dirty region is within a predefined rectangle, 200 by 200 pixels in our implementation, centered at the latest touch point. This also simplifies the implementation. \async will first check if there is any change outside the rectangle around the touch point, i.e., any dirty region outside it. If so, it stops. Otherwise, \pos estimates if the application can finish writing before the display reaches the edge of the rectangle. If yes, it will respond to the application with the \state{busy$_{disp}$} buffer.   

To check if there is a dirty region outside the rectangle, \pos can leverage help from the application, the answer to (1). 
For legacy applications that do not use the partial-drawing APIs, \pos compares the two adjacent frames by sampling. We discuss how we implement it and its overhead in \S~\ref{sec:impl_position} and \ref{sec:overhead}, respectively.

%% file: implementation.tex
\section{Implementation}\label{sec:impl}

\begin{figure}[t]
	\centering
	\includegraphics[width=1\linewidth,trim={2.5cm 3cm 6.5cm 2cm},clip]{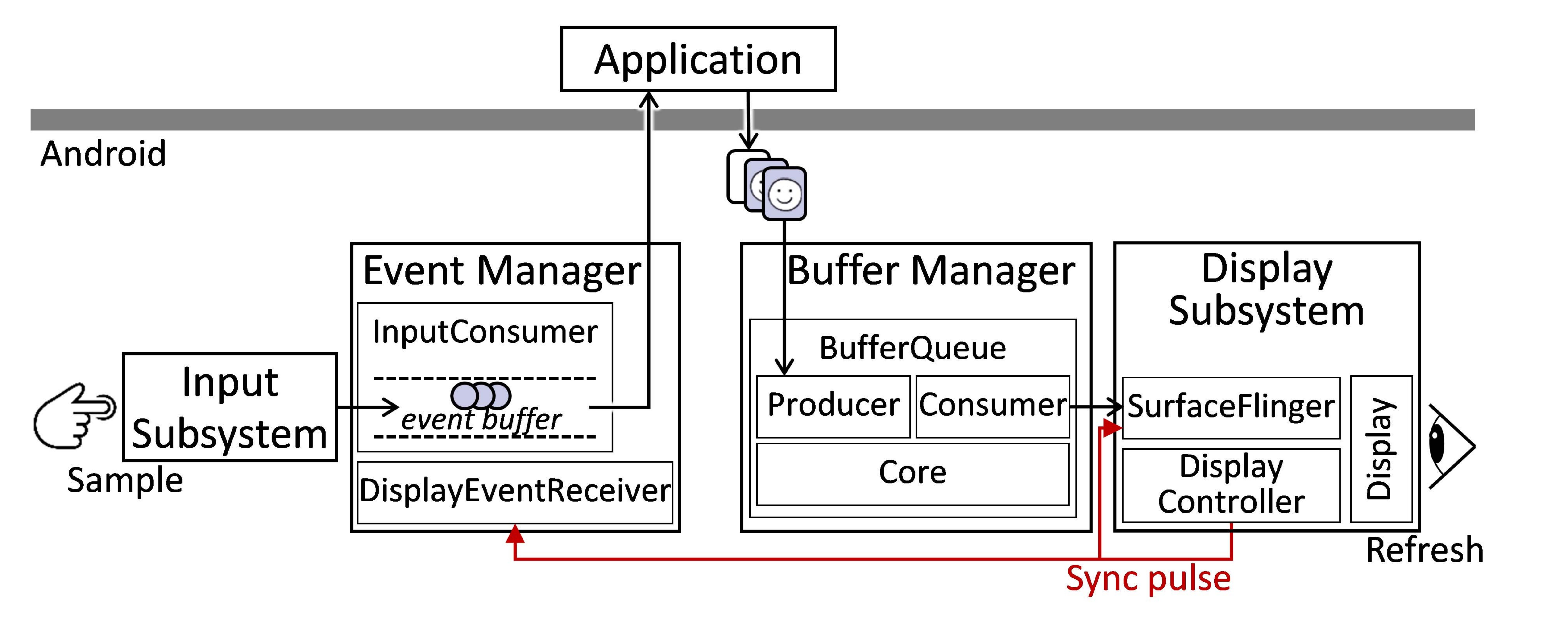}
	\caption{ 
		Android's \route implementation
	}
	\label{fig:impl}
\end{figure}

We first describe Android's I2D path implementation, which is summarized by Figure~\ref{fig:impl}, and then describe our prototype implementation of \poly with two path designs, i.e.,
\async and Android legacy, using Android 5 (Lollipop). 

In Android 5, the event manager includes a library \code{libinput}, and a \vsync receiver \code{DisplayEventReceiver} in Android runtime library \code{libandroid\_runtime}.
Both are in an application's address space.
In the \code{libinput} library, the \code{InputConsumer} object receives events from the input subsystem through a Unix socket and buffers them.
\code{InputConsumer} delivers the buffered events to the application when \code{DisplayEventReceiver} receives a \vsync from the display subsystem.
The buffer manager is \code{BufferQueue}, which is part of Android GUI library \code{libgui} and allocates buffers and manages their ownership. 
Note that we use \code{BufferQueue} to refer to three classes: \code{BufferQueueCore}, \code{BufferQueueProducer}, and \code{BufferQueueConsumer}. \code{BufferQueue} is indirectly synchronized with the \vsync by responding to requests from the display subsystem.
\code{BufferQueue} is not in an application's address space; however, each application window has its dedicated \code{BufferQueue}. As a result, \code{BufferQueue}s from different applications are independent from each other.

The display subsystem includes \code{SurfaceFlinger}, which receives \state{filled} buffers from multiple applications' \code{BufferQueue}s and sets up the hardware composer.
\code{SurfaceFlinger} also relays the \vsync from the hardware composer to the event manager.

\subsection{Implementing \poly}\label{sec:impl_poly}
Implementing \poly involves three major parts: the \manager, \code{ApplyPath()}, and path binding. 
The implementation includes about 740 lines of C++ and 100 lines of Java codes. 

We implement the \manager as an Android system service, which is a thread in the same address space as \code{SurfaceFlinger}.
The event manager, however, is in the application's address space; thus, the \manager communicates with it through a Unix socket named with the application's pid.

We  introduce a new class \code{Polypath} into Android SDK; \code{Polypath} exports the asynchronous path changing API \code{ApplyPath(app\_name,path\_name)}.
When \code{ApplyPath()} is invoked, it sends \code{app\_name} and \code{path\_name} to the \manager through a named Unix socket.
Then, the \manager communicates with the event and buffer managers to apply the path according to the design described in \S\ref{sec:poly}.

To implement paths and binding between path and application, 
we leverage that both data, i.e. events and buffer handles, and control, i.e. \vsync, of Android's \route
are handled by a series of function calls. 
For example, \code{InputConsumer} delivers input events using a chain of function calls, \code{consume()$\rightarrow$consumeBatch()$\rightarrow$consumeSamples()}.
By changing these function calls, we can implement many path designs with various tradeoffs. 
For example, one can add interpolated/predicted events, drop events, or even generate virtual \vsyncs to emulate \SI{120}{Hz} display. We implement \async in the same way.

To allow multiple path designs to coexist, we replace these functional calls with function pointers. 
By default, these pointers point to the corresponding function calls of Android's legacy implementation.
When applying a new path design, we simply redirect the necessary pointers. This implementation is 
efficient in terms of both code size and runtime overhead.
Because each application has its own dedicated event and buffer managers, one obvious alternative is to implement the event and buffer managers for each path and create an instance for each application that asks for the path. For example, we could implement a version of \code{libinput} for each path and an application will simply link to the corresponding version. This mechanism, however, has two disadvantages. First, it adds to software bloat because sharing common code between paths is hard, requiring major refactoring of Android's event and buffer managers. Moreover, it incurs high overhead when an application changes its path, e.g., that of unlinking and linking \code{libinput}.

\subsection{Implementing \async}\label{sec:impl_presto}
\async, as a path the \poly supports, modifies the bindings in the event and buffer mangers for \jit and \pos.
The implementation includes 991 lines of C++, 35 lines of Java, and 212 lines of Linux kernel codes.

\subsubsection{\jit: Just-in-Time Trigger}\label{sec:impl_presto_jitt}
We implement \jit by revising the event manager (\code{libinput} and \code{DisplayEventReceiver}) and buffer manager (\code{BufferQueue}).

The predictor for {\small $T_{app}$} tracks {\small $T_{app}$} history and predicts based on a simple algorithm that averages the recent 32 measurements of {\small $T_{app}$},  or roughly half a second.
We empirically set {\small $T_{out}$} to \SI{3.5}{ms} based on profiling of \code{BufferQueue} and \code{SurfaceFlinger}. The constant time is conservatively determined to give \code{SurfaceFlinger} enough time to transfer the ownership of multiple applications' graphics buffers, from \code{BufferQueue} to the hardware composer.

To trigger the event manager, 
we modified \code{DisplayEventReceiver} to intercept the \vsync from the display subsystem and re-fire it at the predicted time $(t_{refresh} - T')$.
When \code{BufferQueue} is requested to give a \state{filled} buffer by the \code{SurfaceFligner}, it waits until the predicted time $(t_{refresh} - T_{out})$ and then responds with the latest \state{filled} buffer just before the next screen refresh.

\subsubsection{\pos: Position-Aware Buffer Manager}\label{sec:impl_position}
We implement \pos by modifying the buffer manager (\code{BufferQueue}) and Android's ION memory manager in the kernel.
Recall that when the application requests a buffer, \pos responds with the \state{busy$_{disp}$} buffer in the application's buffer manager only if it is confident that the application would finish writing into the buffer BEFORE the display subsystem starts externalizing a dirty region.
Our implementation conveniently obtains the prediction of how long it will take the application to finish writing into the buffer from \jit, i.e., {\small $T'_{app}$}. We profile that the display subsystem reads the \state{busy$_{disp}$}
buffer at \SI{221}{M} pixels per second.

If the application does not already provide information about the dirty region, e.g. via an SDK like~\cite{191579}, our implementation identifies the starting point of the dirty region by modifying Android ION's \code{ioctl()} syscall to compare frames in software. We compare the frames in the kernel space because graphics buffers are not directly accessible from the user space  for security reasons.
\code{BufferQueue} passes a buffer's ION \code{fd} to the kernel via the syscall.
Then, the kernel finds the corresponding memory area represented in \code{scatterlist}~\cite{scatterlist}, samples 1\% of the frame, and then compares them with those of the previous frame.
One can increase the number of samples to track dirty region more accurately; however, 1\% from a 2560$\times$1440 screen (Nexus 6) is sufficient to check the dirty regions of applications updating the entire screen, such as animation and scroll.

%% file: evaluation.tex
\section{Evaluation}\label{sec:eval}

Using the prototype, we seek to answer the following questions regarding \poly and  \async.

\begin{enumerate}
\vspace{-1ex}

\item What is the overhead of \poly?

\vspace{-1.5ex}
\item How effective is \async in reducing latency? how much does each of its two key techniques contribute?

\vspace{-1.5ex}\item Is its effectiveness orthogonal to that of other popular latency reduction technique, namely event prediction~\cite{asano2005predictive, lank2007endpoint, laviola2003double, pasqual2014mouse, iosPrediction}?

\vspace{-1.5ex}\item What tradeoffs does \async make, in terms of power consumption and the visual goals dear to the legacy path design? 

\vspace{-1.5ex}
\item How do users evaluate \async?
\end{enumerate}

\input{eval_method}

\input{eval_latencymeasurement}
\input{eval_overhead}

\input{eval_latency}

\input{eval_user}

%% file: eval_method.tex
\subsection{Evaluation Setup}\label{sec:eval_setup}

We evaluate our implementation on Google Nexus 6 smartphones with Android 5.0 (Lollipop) and Linux kernel 3.10.40.
The smartphone has a 5.96$^{\prime\prime}$ 2560 $\times$ 1440 AMOLED display, 2.7 GHz quad-core CPU, and 600 MHz GPU.
During the evaluation, we use a DotPen stylus pen with a tip of \SI{1.9}{mm}~\cite{dotpen}, instead of finger, to find out the touched position with high accuracy.

\textbf{Benchmarks}:~~
We evaluate \async with both legacy applications and an in-house application. 
Since the effect of the latency is clearer in drawing applications, we select ten drawing applications, the top five each from the \program{Drawing \& Handwriting} and \program{Calligraphy} categories of the Google Play Store on Jan 26, 2016. 
Some of the top applications only provide instructions for calligraphy without drawing facilities; we replaced them by the applications ranked next.
The five from \program{Drawing \& Handwriting} are \program{Notepad+ Free (N+)}, \program{Autodesk Sketch (AD)}, \program{Handrite Note (HN)}, \program{Bamboo Paper (BP)}, and \program{MetaMoJi Note Lite (MM)}.
The five from \program{Callligraphy} are \program{Calligraphy HD (CY)}, \program{Calligrapher (CR)}, \program{INKredible (IK)}, \program{Brush Pen (BP)}, and \program{HandWrite Pro Note (HP)}.
For these applications, we measure the latency using the indirect method presented in \S\ref{sec:measure_indirect}.

We also employ an in-house application because its latency can be directly measured with a more accurate method described in~\S\ref{sec:measure_direct}. The application uses OpenGL ES 2.0 to draw a 115$\times$115 square and a horizontal line on a touched position. 
As the \pen moves, it drags the square and line with it.
We have implemented the application for both Android and iOS and will make both implementations open-source.
The in-house application is valuable for three reasons. (\textit{i}) First, it allows us to understand the accuracy of the indirect measurement of legacy applications. (\textit{ii}) Second, it allows us to compare our Android-based \async prototype with iPad Pro with Apple Pencil, a cutting-edge touch device commercially available, using the same OpenGL ES code base. and \emph{(iii)} Because the application has bare minimum functionality for touch interaction, it allows us to better understand the power overhead of \async.

\textbf{Interaction and Trace Collection}:~~Short of a programmable robotic arm, we try our best to produce repeatable traces of interaction with the benchmarks. For each benchmark, we interact by manually moving the \pen repeatedly from one end of the screen to the other vertically in portrait orientation, with a steady speed for 150 seconds. Post collection analysis shows an average speed of \SI{68}{mm} per second, with a standard deviation of 12. All traces will be made available online.

%% file: eval_latencymeasurement.tex
\subsection{Latency Measurement}
Measuring the end-to-end latency of touch interaction is nontrivial because neither the starting point, the moment of a physical touch, and the end point, the moment of display externalization, can be observed by software running in the mobile device. 
Below we present two measurement methods used in our evaluation of \async. The first one is \emph{indirect}, by combining calibration, analysis, and OS-based time logging. 
It is applicable to all applications. The second is \emph{direct} based on camera capture and video analysis. It is, however, only applicable to applications whose visual effects are amenable to our video analysis. 
In our evaluation, we use the indirect method to report latencies for legacy benchmarks; we use the direct method to provide in-depth insight along with the in-house benchmark.

\subsubsection{Indirect Measurement}\label{sec:measure_indirect}

The indirect measurement method breaks down the end-to-end latency into three parts and deal with each differently: (1) from physical touch to the touch device driver, (2) from the touch device driver to the display subsystem, and (3) from the display subsystem to display externalization. 

We measure the latency of (1) by using a setup with a microcontroller and two light sensors (API PDB-C142, response time: \SI{50}{us}): the microcontroller continuously polls the sensor output at \SI{1}{KHz}. 
We place the first light sensor besides the screen and shoot a laser beam from the other side.
When the stylus pen crosses the laser beam, the light sensor detects it and changes its output; the microcontroller detects the change and logs a timestamp.
When the touch device driver receives an event crossing the beam, it turns on the built-in LED, which takes \SI{1.5}{ms}.
The second light sensor, placed directly above the LED, detects this so that the microcontroller logs the second timestamp. We estimate the latency of (1) as the difference between these timestamps: \SI{28.0}{}$\pm$\SI{1}{ms}.

We measure the latency of (2) by logging two timestamps in software: when the touch device driver receives an interrupt and when the ownership of the resulting buffer is transferred to the display subsystem.
Notably the latency of (2) is where \async makes a difference. 

We estimate the latency of (3) based the y-coordinate of the touch event logged in software as described above.
Since the display panel illuminates pixels sequentially top-down after a \vsync, we estimate when the pixels of the touched area illuminate as {\small $T_{sync}\cdot y/H$} where {\small $H$} is the screen height measured in pixel number.
\subsubsection{Direct Measurement}\label{sec:measure_direct}

For the in-house benchmark, we are able to measure the user-perceived latency by analyzing video record. What a camera can precisely capture are the locations: that of the square ({\small $L_s$}) in response to a touch and that of the \pen ({\small $L_p$}) in each frame.  
Therefore, we estimate the velocity of the \pen movement ({\small $v$}) from its locations in consecutive frames. 
By calculating how long it would take the \pen to travel from the touched location ({\small $L_s$}) to the current \pen location ({\small $L_p$}), we obtain the latency as {\small $(L_s-L_p)/v$}.  
This estimation, however, relies on the assumption that the velocity of the pen does not change abruptly from frame to frame. Due to the high frame rate, i.e., \SI{60}{Hz}, this assumption is largely true and also confirmed by our own measurement.  

We note that the camera also introduces errors due to its limited frame rate.
We use a Nikon D5300 camera with 60 Hz frame rate and 1/500 sec shutter speed.
The frame rate would introduce a random latency uniformly distributed between \SI{0} to \SI{16.7}{ms} ({\small $T_{sync}$}). Therefore, we deduce this random variable when reporting the latency measurement.

We compare the latency derived from the indirect measurement of the in-house application against with its direct measurement for the in-house benchmark with stock Android, \async (\jit) and \async (JITT+\pos).
The direct and indirect measurements are within \SI{2.5}{ms} from each other. The difference is smaller than their standard deviation and more importantly, one order of magnitude smaller than the latency reduction achieved by \async.

%% file: eval_overhead.tex
\subsection{Overhead of \poly}\label{sec:overhead_poly}
We measure the overhead of \poly in terms of application launching delay and path switching delay on the ten drawing benchmarks applications (\S\ref{sec:eval_setup}). 
When an application is launched, \poly imposes the extra overhead to acquire the application's path preference from \code{PackageManager} and to bind the path to the application. 
Our measurement shows that the extra delay is between \SI{13.1}{} to \SI{34.5}{ms}, which is negligible for the launch delays of many \SI{100}{ms} for legacy applications.

The path changing API \code{ApplyPath()} is asynchronous in order to ensure the path integrity (\S\ref{sec:poly}). 
There is a delay between when \code{ApplyPath()} is invoked and when the new path is in place. 
Note this delay is not part of the latency of the \route.
We estimate the delay to be between \SI{0}{ms} and {\small $3\cdot T_{sync}$}.
The worst delay happens when the path changes from the legacy design to \async and \code{ApplyPath()} is called when the event manager has only one event in its buffer.
In this case, the event manager will wait almost {\small$T_{sync}$} to buffer more events and deliver them to the application, and the buffer manager 
will wait another {\small$2\cdot T_{sync}$} to externalize two frames, one that the application is currently drawing and the other resulting from the buffered events.
We measured the time from when the API is called to when the buffer manager finishes the path change, which is the completion of the path change procedure (\S\ref{sec:poly}).
Our measurements confirmed the above estimation with 1000 path changes each when the \route is active and inactive, respectively.
When the \route is active, the average and standard deviation are \SI{20.5}{ms} and \SI{10.1}{ms}, respectively; when inactive, they are \SI{0.27}{ms} and \SI{0.08}{ms}, respectively.

%% file: eval_latency.tex
\subsection{Latency Tradeoff by \async}
We next answer the three questions about the latency reduction, its orthogonality and the tradeoffs by \async.
The measurement shows that \async with \jit only and with \jit and \pos reduces the average latency of our benchmarks from \SI{72.7}{ms} to \SI{54.4}{ms} and \SI{41.0}{ms}, respectively. 
This reduction eliminates all latency from synchronization and will significantly improve user experience and performance according to both the literature~\cite{deber2015much} and our own experience and user study.
Moreover, as we anticipated, the reduction from \async is orthogonal from that of another important technique, \tp, employed by iPad Pro. When combined with \tp of \SI{30}{ms}, \async is able to reduce the latency of our in-house application below \SI{10}{ms}.

\subsubsection{\async reduces latency by \SI{32}{ms}} 

\begin{figure}[t]
		\centering
		\includegraphics[width=1\linewidth]{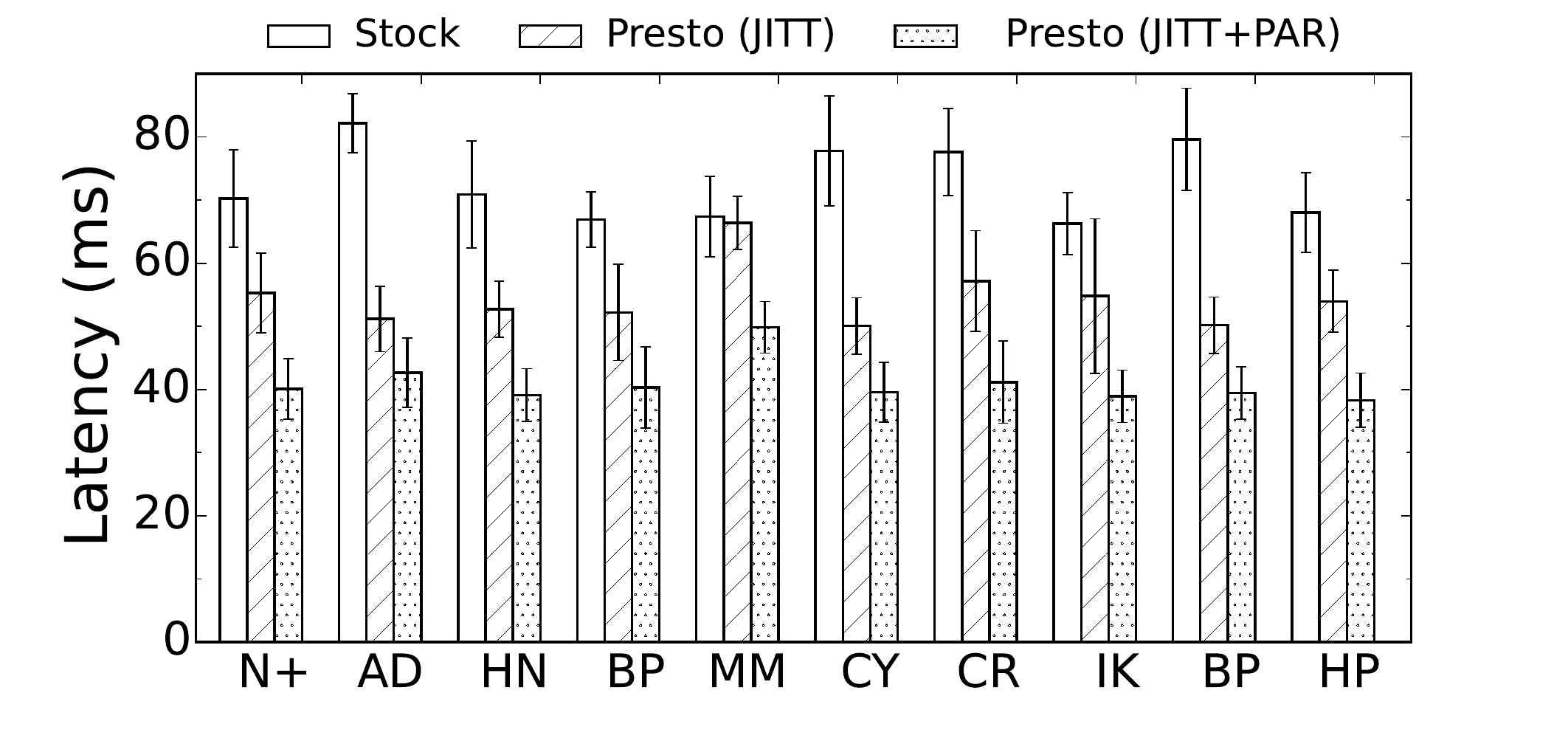}
		\caption{\async consistently reduces the latency of the legacy benchmarks, by \SI{32}{ms} on average.}
		\label{fig:latency_reduction}
\end{figure}

Figure~\ref{fig:latency_reduction} shows how much each of the two techniques reduces the latencies of legacy applications.
On average, \async reduces the latency by \SI{32}{ms}. 
To appreciate the significance of this reduction, we note that Deber et al~\cite{deber2015much} showed that even a small latency reduction, i.e. \SI{8.3}{ms}, brings a perceptible effect in touchscreen interactions.
This reduction is larger than average latency caused by the coarse-grained \route design, i.e., \SI{26.6}{ms} (\S\ref{sec:monopath_latency}).
This is because when an application occasionally fails to finish drawing by the next display refresh, \jit drops this delayed frame while stock Android keeps it and propagates the delay to all subsequent frames.
The frame drop by \async is not perceptible to users as we will see in\S\ref{sec:user_eval}.

Notably, different benchmarks see different amount of latency reduction from \async. 
\async is most effective for those that have large latency to begin with, i.e., \program{Autodesk (AD)}, \program{Calligraphy (CY)}, \program{Calligrapher (CR)}, and \program{Brush (BP)}.
\async is the least effective for \program{MetaMoji (MM)}, reducing the latency by \SI{17.6}{ms} only. 
Our analysis reveals that this is because its average {\small $T_{app}$} is the longest among all benchmarks. As a result,
it has the least amount of latency due to synchronization and gives \async the least opportunity.

\subsubsection{\async beats iPad Pro}

\begin{figure*}[t]
	\begin{minipage}[t]{0.48\textwidth}
		\centering
		\includegraphics[width=1\linewidth,trim={1.4cm 0 2.4cm 0cm},clip]{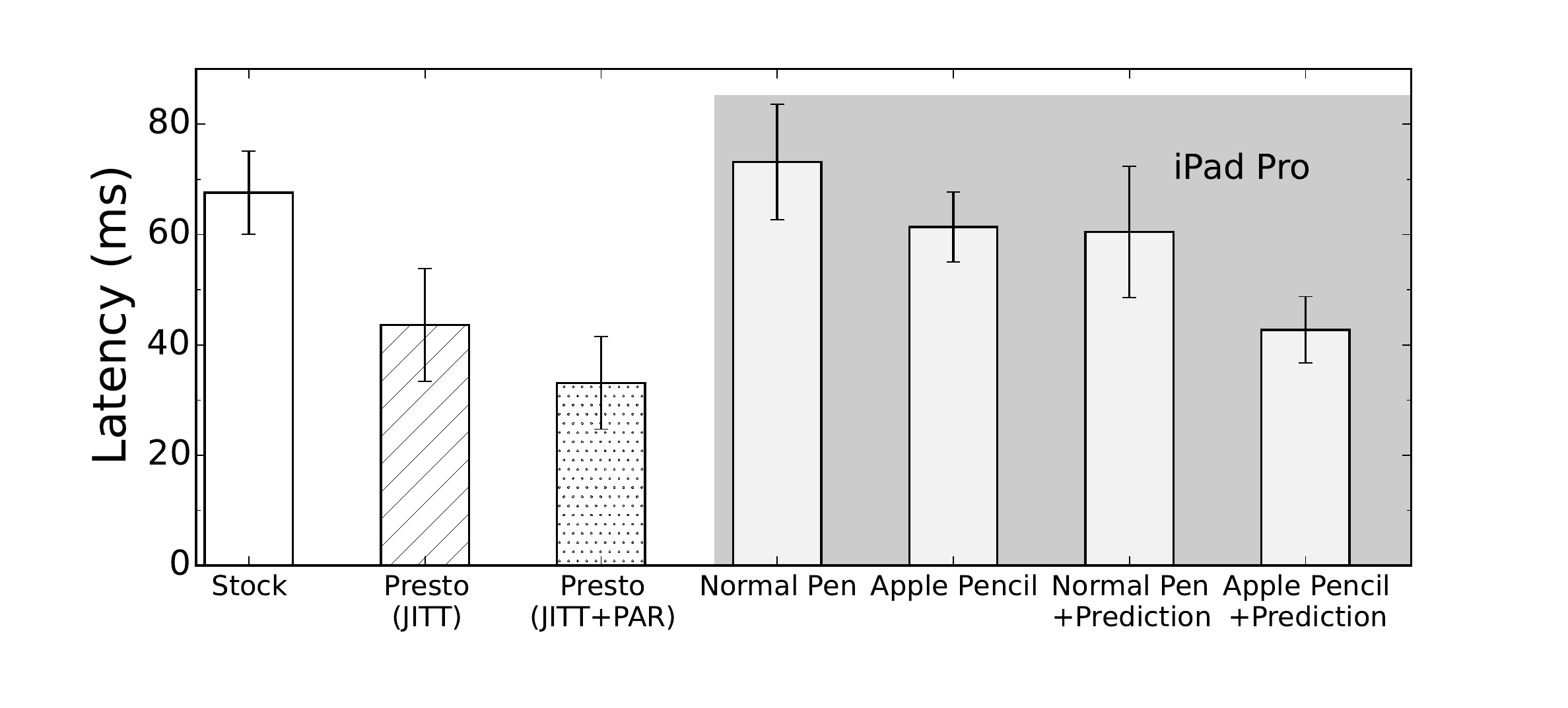}
		\caption{\async improves the latency of our in-house application on Nexus 6 below that of iPad Pro even with Apple Pencil and \tp. 
		}
		\label{fig:latency_ipadPro}
	\end{minipage}
	\hfill
	\begin{minipage}[t]{0.48\textwidth}
		\centering
		\includegraphics[width=1\linewidth]{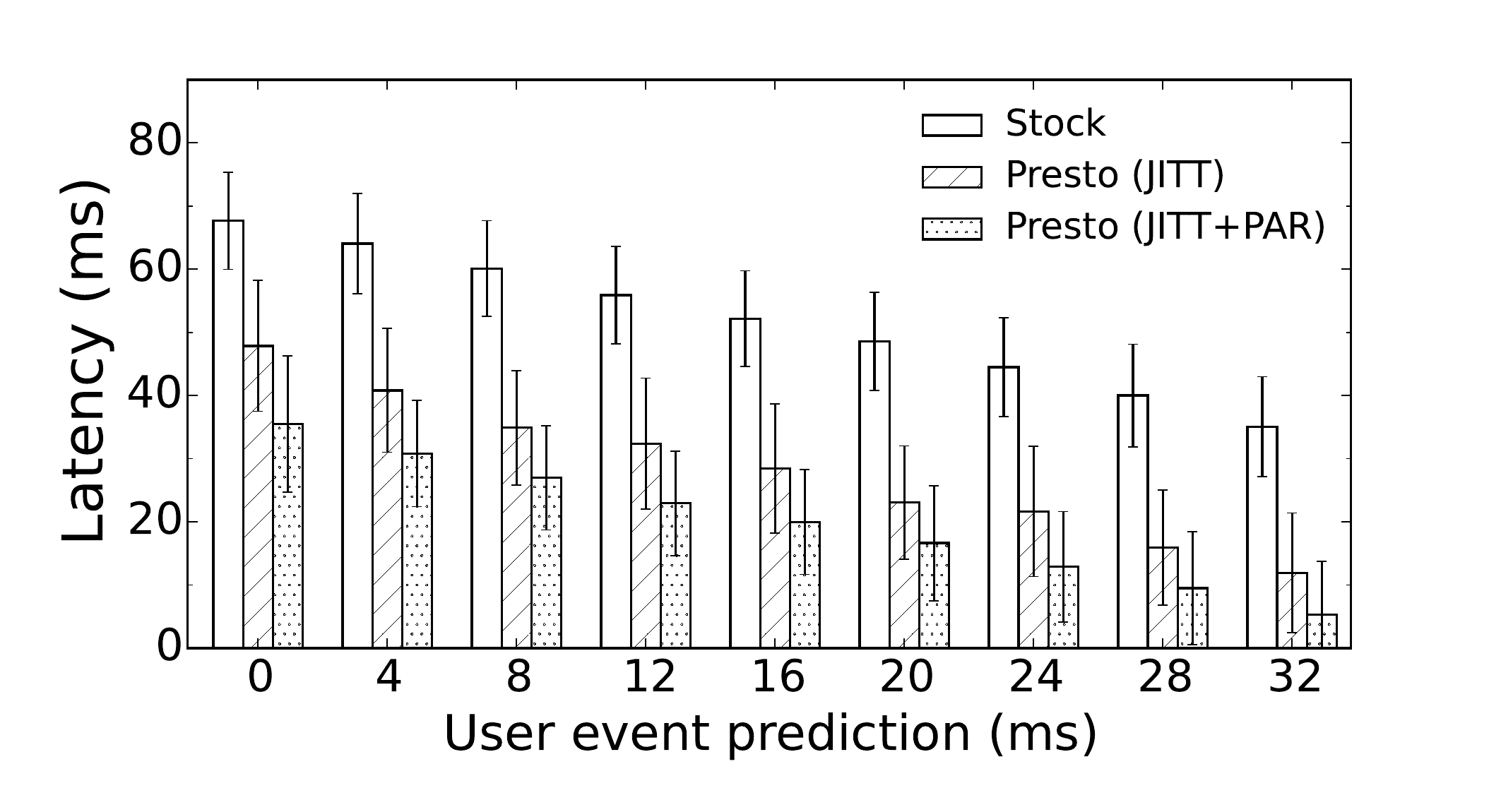}
		\caption{Latency of \async plus \tp for our in-house application: the effectiveness of \async is complementary to that of \tp. X axis is the time into the future predicted. 
		}
		\label{fig:latency_result}
	\end{minipage}
\end{figure*}

Using our in-house application, we are able to compare \async on Nexus 6 with iOS on iPad Pro, the state-of-the-art touch device widely in use. 
iPad Pro employs two techniques to reduce the latency: it doubles the input sampling rate for Apple Pencil~\cite{applePencil}, from 120 Hz to 240 Hz, and the iOS SDK provides predicted events for the next frame (\SI{16}{ms}), a technique called \emph{\tp}~\cite{iosPrediction}. 
Because neither technique is available on Android, we measure the in-house application on iPad Pro with four configurations as reported in the right half of Figure~\ref{fig:latency_ipadPro}: normal stylus pen without \tp, Apple Pencil without \tp, normal stylus pen with \tp, and Apple Pencil with \tp.  
The results clearly show that both the faster input sampling rate and \tp help reduce the latency for iPad Pro, with the best latency being \SI{42.9}{ms}. Impressively, \async is able to reduce Android's latency to even lower, \SI{33.0}{ms}, even without fast input sampling or \tp. 

\subsubsection{\async brings orthogonal benefits}\label{sec:orthogonal}

In principle, the effectiveness of \async is orthogonal to that of faster input and \tp 
because \async eliminates latency resulting from synchronization and the latter primarily reduce latency resulting
from the input hardware.
With our in-house application, we implement \tp that predicts into the future from \SI{0} to \SI{32}{ms}.
\async reduces the latency by eliminating the synchronization points.
Figure~\ref{fig:latency_result} shows how \async and \tp complementarily reduces the latency.
The leftmost group in the figure does not have predicted events, i.e., \tp of \SI{0}{ms}.
Clearly, for \tp of various time, \async demonstrates almost the same effectiveness in latency reduction.
Interestingly, \async with \tp of about \SI{30}{ms} is able to reduce the average latency below \SI{10}{ms}, a rather 
remarkable achievement by a software-only solution.

\subsubsection{Tradeoffs by \textbf{\async}}\label{sec:overhead}
\async trades off other computing goals for short latency: it judiciously allows frame drops and tearing, and may incur power overhead through \pos.
When we try out the benchmarks with \async, we could not see any effects usually associated with frame drops or tearing.
Our double-blind user study, reported in \S\ref{sec:user_eval}, confirms this independently. Below we report objective data regarding frame drops, tearing risk, and power overhead.

By design, \async guarantees no consecutive frame drops.  In the worst case, it would drop \SI{50}{\%} of the frames (every other frame). Our measurement, reported in Figure~\ref{fig:frameDropTearing},
shows a much lower rate for our benchmarks, with the worst case being \SI{8}{\%} (\program{Bamboo (BP)}).

There is no direct way we could observe the occurrences of tearing: as shown in \S\ref{sec:newrequirements}, even if tearing happens and is captured by camera, it would be extremely hard to tell it from the effect of latency.
Instead, we measure how frequent underprediction of {\small $T_{app}$} happens. As shown in \S\ref{sec:design_position}, an underprediction of {\small $T_{app}$} is a necessary but not sufficient condition for tearing to happen. Therefore, the frequency of underprediction can be considered as an upper bound for that of tearing.
 Figure~\ref{fig:frameDropTearing} shows the frequencies of underprediction for the legacy benchmarks. \program{HandWrite (HP)} has the highest frequency of underprediction (\SI{17}{\%}). \program{Bamboo (BP)} has the highest frequency (\SI{13}{\%}) amongst the five benchmarks used in the user study. These frequencies are at most suggestive of how often tearing may happen. None of the authors could see any effects due to tearing; nor did the participants in our user study.

We use a Monsoon Power Monitor~\cite{monsoon} to measure the power consumption of \async in Nexus 6.
We disable all wireless communications and dim the LCD backlight to the minimum level.
We measure the power consumption of the in-house application during 60 seconds of touchscreen drawing for each of the following configurations:
 without \async, with \async (\jit), \async (JITT+\pos without frame comparison), and \async (JITT+\pos). 
 Their power consumption and standard deviations are $2017 \pm 120$, $2075\pm115$, $2024 \pm 110$, and $2564 \pm 201$ \SI{}{mW}, respectively. 

 We would like to highlight two points regarding the power overhead.
First, \jit increases the power consumption only slightly, well below the standard deviation. \pos (without frame comparison) decreases the power consumption to be barely indistinguishable from that of the stock Nexus 6, i.e. $2024 \pm 110$ vs. $2017 \pm 120$. This is because \pos reduces activities of the buffer manager. 
Second, the frame comparison needed for \pos contributes most of the power overhead, an 27\% increase. Because in our measurement we disabled all wireless interfaces and dimmed the LCD backlight to minimum, the percentage increase for real-world usage will be much lower. More importantly, using frame comparison to determine the dirty region is not practically necessary because the GPU and application already have the information. Some SDKs, e.g.,\cite{glsurfaceview}, already make this information available via an API, e.g., \code{invalidate(Rect dirty)}. With such APIs, this overhead would be eliminated.

%% file: eval_user.tex
\subsection{User Evaluation}\label{sec:user_eval}

\begin{figure*}[t]
	\begin{minipage}[t]{0.48\textwidth}
		\vspace{0pt}
		\centering
		\includegraphics[width=1\linewidth]{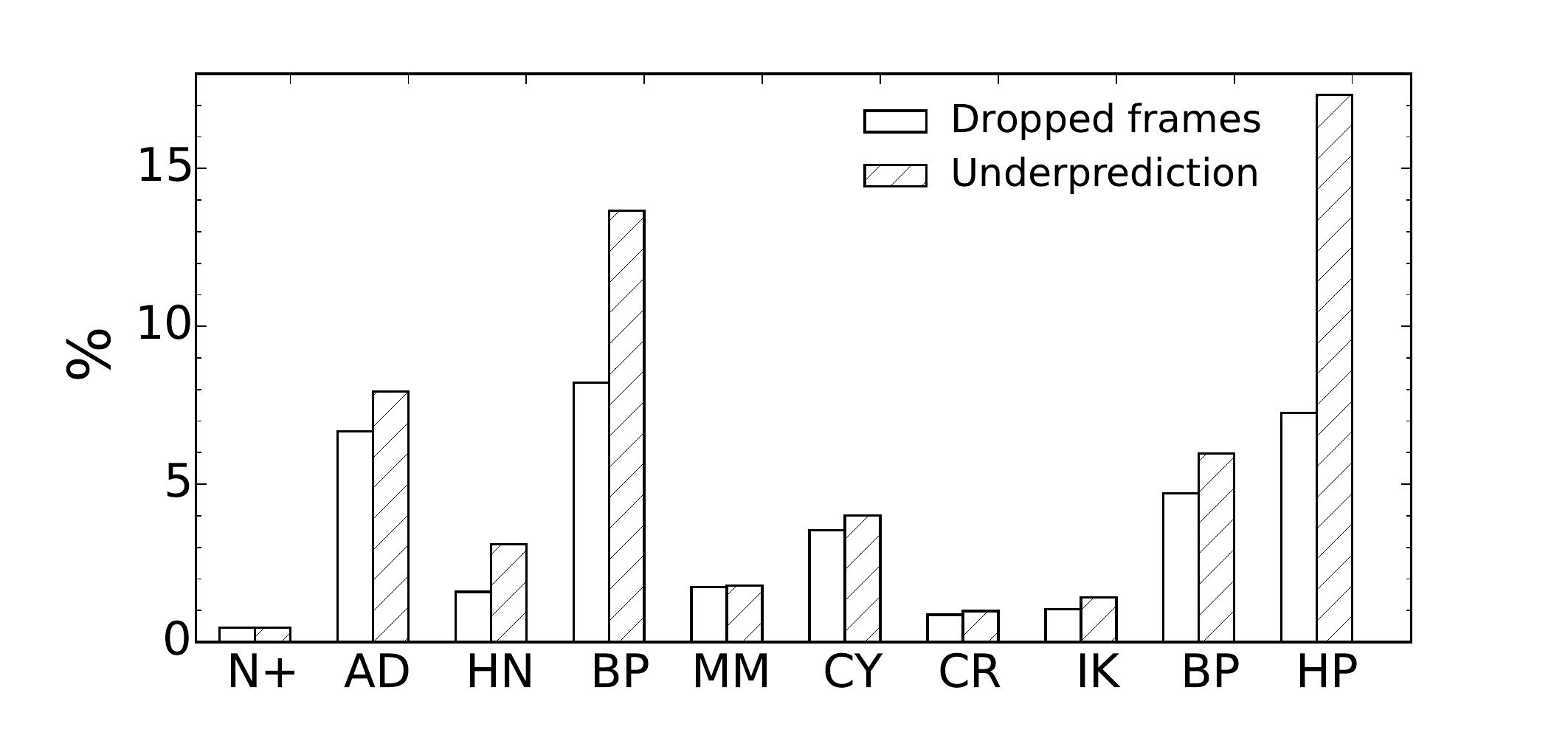}
		\caption{\async occasionally experiences frame drops and underprediction. 
		}
		\label{fig:frameDropTearing}
	\end{minipage}
	\hfill
	\begin{minipage}[t]{0.48\textwidth}
		\vspace{0pt}
		\centering
		\includegraphics[width=1\linewidth]{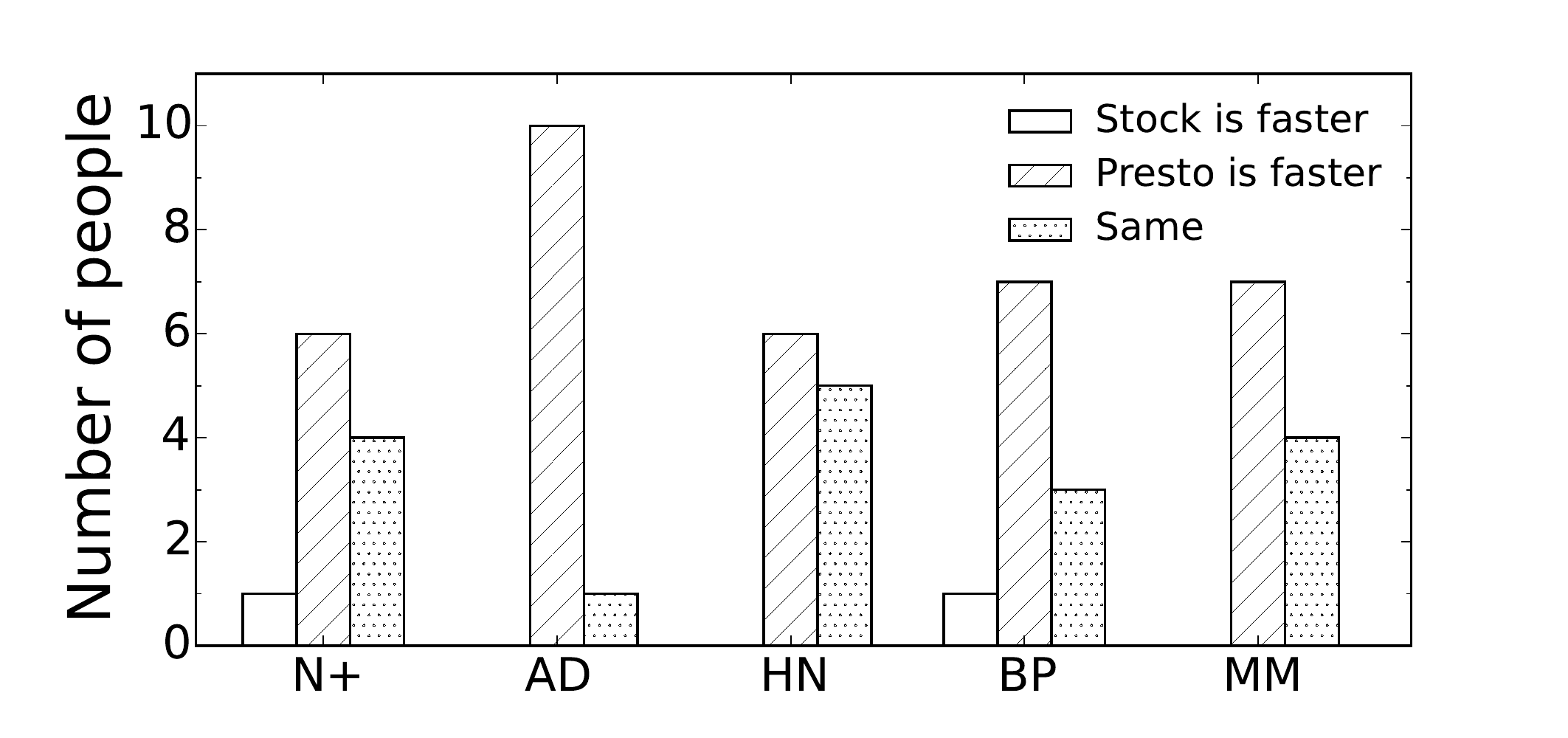}
		\caption{Number of participants answering Question (\textit{i}) in each of the three ways: which device is faster: A, B or same? 
		}
		\label{fig:userstudy}
	\end{minipage}
\end{figure*}

When we try the benchmarks with \async, it is visually obvious that \async reduces latency significantly. None of us are able to notice any tearing effects or frame drops. Nevertheless, In defense against any possible experimenter's bias, we perform a double-blind user study to evaluate \async subjectively. 

\textbf{Participants}:~We recruited 11 participants via campus-wide flyers. They were students and staff members from various science and engineering departments, between 19 and 40 years old, with three women. 
All had at least two-month experience with an Android device with a display bigger than 5.5 inches.

\textbf{Procedure}:~Each participant came to the lab by appointment and was given two Nexus 6 smartphones that are identical except one has stock Android, the other \async. The smartphones are marked A and B, respectively. Neither the participant nor the study administrator knew which one is stock. The participant was then asked to use their finger or a stylus pen to try out the five top Android applications from \program{Drawing \& Handwriting}. They were allowed to try as long as they wished; and all finished in 10 to 45 minutes. After each application, the participant answered three questions: (i) which device is faster: A, B or same? (ii) if you chose A or B, to what extent do you agree with the statement that the latency difference is obvious? (1 to 5 with 1 being \emph{strongly disagree} and 5 being \emph{strongly agree})
(iii) other than the latency, describe any difference you observe.
For post-mortem analysis, we recorded the hand-smartphone interaction of all except two participants with a GoPro Hero 4 camera at \SI{240}{Hz}. We plan to release the video clips in compliance with our institutional review board (IRB) approval.

\textbf{Quantitative findings}:~Figure~\ref{fig:userstudy} presents participants' answers to the first question.
Not surprisingly, more than half of the participants consider \async to be faster in each of the benchmarks.\footnote{We emphasize that the participants do not know which device has \async. The identity is only used in our data analysis and presentation.}
For \program{Autodesk (AD)}, 10 out of 11 participants considers \async is faster. This corroborates the measurement presented in Figure~\ref{fig:latency_reduction}, which shows \program{Audtodesk (AD)} sees the largest latency reduction amongst the five.
To our surprise and puzzlement, a same participant reported the stock Android is faster in \program{Notepad+ (N+)} and \program{Bamboo (BP)}.
We checked the video record, and it was obvious to us that \async was clearly faster in both the applications. 
One theory to explain this is that the participant mistook A with B when answering the question. Nevertheless, we are wary that the same theory can be used to argue the participant's responses for the other three applications were also mistaken. Overall, the data suggests that participants overwhelmingly felt that \async is faster. For those who considered \async to be faster, the average of their responses to the second question is 3.5, indicating the latency difference is obvious to them.

\textbf{Qualitative findings}:~Our participants were asked if they observe any difference beyond latency. None reported any effects that may result from inconsistent frame rate, frame drop, or tearing, such as application's fluctuating response time, screen flickering, and screen overlap.
Indeed most of their comments are about secondary effects due to latency difference. 
Two participants did notice some details about how \async actually works. One remarked about \program{MetaMoji (MM)} that \async ``seems to catch up quicker than'' the stock Android. The other observed similar effects with \program{Autodesk (AD)} but worded it differently: the stock Android has ``smooth curves;'' \async is ``not as soft as'' the stock Android. By that, the participant was referring to the same effect that when drawing a line, the line with \async sometimes jumps to the touch point, or ``catch up quicker'' in the words of the first participant.

%% file: related.tex
\section{Related Work}\label{sec:related}

There is a large body of literature from the systems community that reduces interaction latency by proper OS design. 
To the best of our knowledge, \poly is the first in the public domain that serves different \route designs to different interactions;
\async is the first \route design that achieves low latency by eliminating synchronization and buffer atomicity in \route.

\textbf{Resource Management for Low Latency}:~~A faster computer system reduces the application execution time {\small$(T_{app}+T_{out})$} (\S\ref{sec:monopath_latency}). The OS can also favor interactive applications in resource management to reduce their {\small$(T_{app}+T_{out})$}~\cite{endo1996using,jones1997cpu,endo2000improving,goel2002supporting,yan2005mascots,yang2008redline}.
Many have explored the use of cloud resources to improve the interactive performance of mobile applications, e.g.,~\cite{gordon2012osdi,gordon2015mobisys}.
These solutions are complementary to \async: they reduce latency when {\small$(T_{app}+T_{out})>T_{sync}$} while \async is most effective when {\small$(T_{app}+T_{out})<T_{sync}$}. Additionally, when {\small$(T_{app}+T_{out})<T_{sync}$}, these solutions improve the opportunity for \async by reducing {\small$T_{app}+T_{out}$} as in Equation~\ref{eq:latency}. 

\textbf{Speculation for Low Latency}:~~Event prediction and speculative execution have also been studied to conceal latency.
Event prediction, or \tp in Apple's term, is widely used for the virtual reality with the head mounded display.
To compensate for prediction errors, researchers have explored speculative execution~\cite{lee2015outatime} and post image processing~\cite{mark1997post}. All these solutions, as discussed in~\S\ref{sec:orthogonal}, are complementary to \async, and can be implemented as their own paths in \poly.

\textbf{Specialized Hardware for Low Latency}:~~As part of a testbed for studying touch latency, Ng et al. report an ultra-low latency touch system~\cite{ng2012designing,ng2014blink} that achieves  a latency as short as \SI{1}{ms}. 
The system employs a proprietary touch sensor with a very high sampling rate (\SI{1}{KHz}), FPGA-based
low-latency (\SI{0.1}{ms}) data processing, and an ultra-high speed digital light projector (\SI{32000}{fps}).
With completely custom software and hardware, it is not feasible for mobile systems, let alone supporting any legacy applications as \async does.

\textbf{Alternatives to VSync}:~~Games on non-mobile devices often provide an asynchronous, or \emph{vsync-off}, mode to reduce latency.
In the vsync-off mode, the event manager delivers input events to the game without any delay. The game processes events without waiting for a \vsync; when the game is in the middle of processing events, it buffers the events. Similarly, the buffer manager swaps graphics buffers without waiting for a \vsync, even when the display is reading.
This vsync-off mode, unfortunately, can introduce tearing effects anywhere on the screen~\cite{fastsync}.
\jit avoids it by swapping graphics buffers only when a \vsync is fired; \pos checks dirty regions and confines the tearing effects, if any, to a small area under the touch position.
In contrast, the vsync-off mode blindly ignores the \vsyncs.
Furthermore, simply disabling vsync on mobile devices is not feasible because display controllers on SoCs~\cite{instrumentsomap4460,tegrak1} swap buffers only at a \vsync regardless of when an application finishes rendering.
Suppose that an application generates a frame \SI{1}{ms} before a \vsync and another application generates a frame \SI{5}{ms} before a \vsync.
Regardless of when the rendering finishes, the two frames will be swapped and displayed at the next \vsync.
When a display controller swaps buffers only on a \vsync,
an application should generate multiple frames within a \vsync period in order to reduce latency.
Generating multiple frames within a \vsync period leads to higher GPU power consumption on a mobile device~\cite{bui2015rethinking,nixon2014mobile,yan2015optimizing} as well as more memory usage.

Nvidia's G-Sync~\cite{g-sync} reduces latency in a way very similar to \jit but requires proprietary GPU and display. \jit times the event manager carefully so that the resulting frame will be ready to display right before the next \vsync. In contrast, a G-Sync GPU generates a \vsync when it finishes rendering to synchronize the event and buffer managers, and the display. 

\textbf{Latencies in VR Systems}:~~State-of-the-art tethered VR systems~\cite{htcVive,oculusRift} have latencies between \SI{20}{} and \SI{22}{ms}, much lower than those on mobile systems.
However, the tethered systems are very different from mobile systems in hardware and software.
They have much faster and more power-hungry hardware: inertial sensors with high sampling rates (e.g., \SI{1}{KHz}~\cite{lavalle2014head}) and low latencies (e.g., \SI{2}{ms}~\cite{oculusSensor}), powerful GPUs, and higher display refresh rates (e.g., \SI{90}{} or \SI{120}{Hz}).
Their software takes away the composer and remove one $T_{sync}$ period from the \route because VR systems has only one foreground application at a time.

Importantly, indirect input devices used in VR controllers make users less sensitive to the latency~\cite{deber2015much}.
However, direct input devices such as see-through displays used in AR systems~\cite{hololens,moverio} and touchscreens, which this paper focuses on, manifest latency as a spatial gap and require a lower latency.
\async with \poly may reduce latency from AR and VR systems; however, the latency reduction will be ineffective because the average latency due to the coarse-granularities is smaller than \eqref{eq:latency} and the latency caused by the input hardware is much smaller, too.

Finally, the focus of this work, the \route, is reminiscent of the \emph{path} abstraction that represents data and control flows crossing layered architectures~\cite{hutchinson1991x,przemyslaw1995extensibility,mosberger1996making,ford1997flux,kohler2000click,fassino2002think}. 
Despite the conceptual similarity, the \route is 
unique in its system context, its periodical pace and its concern with latency rather than throughout.

%% file: discussion.tex
\section{Concluding Remarks}\label{sec:discussion}
In this work, we present \poly, an operating system design that supports multiple tradeoffs for interaction latency, and \async, an \route design that halves the latency by judiciously allowing frame drops.
\poly exports an asynchronous API that allows an unmodified legacy application to changes its path design with the guaranteed path integrity, independently from other applications.

\async is able to reduce the average latency of the drawing benchmarks tested from about \SI{70}{ms} to \SI{40}{ms}.
\emph{Where does the rest of latency come from?} Our investigation has pointed to the input hardware, which contributes about \SI{30}{ms} in state-of-art Android systems.
This includes the hardware time for scanning capacitance changes on the touch sensor, converting analog signals to digital, and communicating to the CPU~\cite{touchproduct}.
This latency can be reduced in two ways. One, exemplified by Apple Pencil, is to increase the input sampling rate, as shown in Figure~\ref{fig:latency_ipadPro}.
The more effective way, however, is \tp, as exemplified by iOS 9, as shown in Figure~\ref{fig:latency_ipadPro} and Figure~\ref{fig:latency_result}.

\section*{ACKNOWLEDGEMENTS}
\noindent This work was supported in part by NSF Award  CNS \#1422312.

%% file: main.bbl
\begin{thebibliography}{10}

\bibitem{applePencil}
Apple.
\newblock Apple pencil.
\newblock \\http://www.apple.com/apple-pencil/, 2015.

\bibitem{asano2005predictive}
T.~Asano, E.~Sharlin, Y.~Kitamura, K.~Takashima, and F.~Kishino.
\newblock Predictive interaction using the delphian desktop.
\newblock In {\em Proc. Ann. ACM Symp. User Interface Software \& Technology
  (UIST)}, 2005.

\bibitem{przemyslaw1995extensibility}
B.~N. Bershad, S.~Savage, P.~Pardyak, E.~G. Sirer, M.~E. Fiuczynski, D.~Becker,
  C.~Chambers, and S.~Eggers.
\newblock Extensibility, safety and performance in the {SPIN} operating system.
\newblock In {\em Proc. ACM Symp. Operating Systems Principles (SOSP)}, 1995.

\bibitem{scatterlist}
J.~E. Bottomley.
\newblock Dynamic dma mapping using the generic device.
\newblock https://www.kernel.org/doc/ Documentation/DMA-API.txt.

\bibitem{bui2015rethinking}
D.~H. Bui, Y.~Liu, H.~Kim, I.~Shin, and F.~Zhao.
\newblock Rethinking energy-performance trade-off in mobile web page loading.
\newblock In {\em Proceedings of the 21st Annual International Conference on
  Mobile Computing and Networking}, pages 14--26. ACM, 2015.

\bibitem{claypool2007frame}
K.~T. Claypool and M.~Claypool.
\newblock On frame rate and player performance in first person shooter games.
\newblock {\em Multimedia systems}, 2007.

\bibitem{Claypool:2006:LPA:1167838.1167860}
M.~Claypool and K.~Claypool.
\newblock Latency and player actions in online games.
\newblock {\em Commun. ACM}, Nov. 2006.

\bibitem{Claypool:2010:LKP:1730836.1730863}
M.~Claypool and K.~Claypool.
\newblock Latency can kill: Precision and deadline in online games.
\newblock In {\em Proc. Ann. ACM Int. Conf. Multimedia Systems}. ACM, 2010.

\bibitem{claypool2006effects}
M.~Claypool, K.~Claypool, and F.~Damaa.
\newblock The effects of frame rate and resolution on users playing first
  person shooter games.
\newblock In {\em Electronic Imaging}. Int. Society for Optics and Photonics,
  2006.

\bibitem{deber2015much}
J.~Deber, R.~Jota, C.~Forlines, and D.~Wigdor.
\newblock How much faster is fast enough?: User perception of latency \&
  latency improvements in direct and indirect touch.
\newblock In {\em Proc. ACM Conf. Human Factors in Computing Systems (CHI)},
  2015.

\bibitem{agawi}
D.~E. Dilger.
\newblock Agawi touchmark contrasts ipad's fast screen response to laggy
  android tablets.
\newblock
  \\http://appleinsider.com/articles/13/10/08/agawi-touchmark-contrasts-ipads-fast-screen-response-to-laggy-android-tablets,
  2013.

\bibitem{dotpen}
Dot-Tec.
\newblock Dot pen.
\newblock http://dot-tec.com.

\bibitem{endo2000improving}
Y.~Endo and M.~Seltzer.
\newblock Improving interactive performance using {TIPME}.
\newblock In {\em Proc. ACM SIGMETRICS}, 2000.

\bibitem{endo1996using}
Y.~Endo, Z.~Wang, J.~B. Chen, and M.~I. Seltzer.
\newblock Using latency to evaluate interactive system performance.
\newblock In {\em Proc. USENIX Conf. Operating Systems Design \& Implementation
  (OSDI)}, 1996.

\bibitem{moverio}
Epson.
\newblock Moverio.
\newblock \url{http://www.epson.com/Moverio}.

\bibitem{fassino2002think}
J.-P. Fassino, J.-B. Stefani, J.~L. Lawall, and G.~Muller.
\newblock Think: A software framework for component-based operating system
  kernels.
\newblock In {\em Proc. USENIX Annual Technical Conference (ATC)}, 2002.

\bibitem{ford1997flux}
B.~Ford, G.~Back, G.~Benson, J.~Lepreau, A.~Lin, and O.~Shivers.
\newblock The {Flux OSKit}: A substrate for kernel and language research.
\newblock In {\em Proc. ACM Symp. Operating Systems Principles (SOSP)}, 1997.

\bibitem{goel2002supporting}
A.~Goel, L.~Abeni, C.~Krasic, J.~Snow, and J.~Walpole.
\newblock Supporting time-sensitive applications on a commodity {OS}.
\newblock In {\em Proc. USENIX Conf. Operating Systems Design \& Implementation
  (OSDI)}, 2002.

\bibitem{glsurfaceview}
Google.
\newblock Glsurfaceview.
\newblock \\ http://developer.android.com/reference/android/
  opengl/GLSurfaceView.html.

\bibitem{graphicxArch}
Google.
\newblock Graphics architecture.
\newblock \\ http://source.android.com/devices/graphics/ \\ architecture.html.

\bibitem{graphicxImpl}
Google.
\newblock Implementing graphics.
\newblock http://source.and\\roid.com/devices/graphics/implement.html.

\bibitem{gordon2015mobisys}
M.~S. Gordon, D.~K. Hong, P.~M. Chen, J.~Flinn, S.~Mahlke, and Z.~M. Mao.
\newblock Accelerating mobile applications through {Flip-Flop} replication.
\newblock In {\em Proc. ACM Int. Conf. Mobile Systems, Applications, \&
  Services (MobiSys)}, 2015.

\bibitem{gordon2012osdi}
M.~S. Gordon, D.~A. Jamshidi, S.~Mahlke, Z.~M. Mao, and X.~Chen.
\newblock {COMET}: Code offload by migrating execution transparently.
\newblock In {\em Proc. USENIX Conf. Operating Systems Design \& Implementation
  (OSDI)}, pages 93--106, 2012.

\bibitem{191579}
M.~Ham, I.~Dae, and C.~Choi.
\newblock {LPD}: Low power display mechanism for mobile and wearable devices.
\newblock In {\em Proc. USENIX Annual Technical Conference (ATC)}, 2015.

\bibitem{henderson1976lazy}
P.~Henderson and J.~H. Morris~Jr.
\newblock A lazy evaluator.
\newblock In {\em Proc. ACM SIGACT-SIGPLAN Symp. Principles on Programming
  Languages (POPL)}, 1976.

\bibitem{htcVive}
HTC.
\newblock Htc vive.
\newblock \url{https://www.htcvive.com}.

\bibitem{hutchinson1991x}
N.~C. Hutchinson and L.~L. Peterson.
\newblock The x-kernel: An architecture for implementing network protocols.
\newblock {\em IEEE Trans. Software Engineering}, 1991.

\bibitem{janzen201460}
B.~F. Janzen and R.~J. Teather.
\newblock Is 60 fps better than 30?: The impact of frame rate and latency on
  moving target selection.
\newblock In {\em Proc. ACM Conf. Human Factors in Computing Systems (CHI)},
  2014.

\bibitem{lesnumeriques_21devices}
R.~Jehl.
\newblock Le retard tactile de l'\'ecran de 21 smartphones et tablettes.
\newblock
  \\http://www.lesnumeriques.com/telephone-portable/reactivite-tactile-ecran-21-smartphones-tablettes-n29229.html,
  2013.

\bibitem{jones1997cpu}
M.~B. Jones, D.~Ro{\c{s}}u, and M.-C. Ro{\c{s}}u.
\newblock {CPU} reservations and time constraints: Efficient, predictable
  scheduling of independent activities.
\newblock In {\em Proc. ACM Symp. Operating Systems Principles (SOSP)}, 1997.

\bibitem{kohler2000click}
E.~Kohler, R.~Morris, B.~Chen, J.~Jannotti, and M.~F. Kaashoek.
\newblock The click modular router.
\newblock {\em ACM Transactions on Computer Systems (TOCS)}, 2000.

\bibitem{touchproduct}
S.~Kolokowsky and T.~Davis.
\newblock {\em Not All Touchscreens are Created Equal - How to ensure you are
  developing a world class touch product}.
\newblock Planet Analog, 2010.

\bibitem{lank2007endpoint}
E.~Lank, Y.-C.~N. Cheng, and J.~Ruiz.
\newblock Endpoint prediction using motion kinematics.
\newblock In {\em Proc. ACM Conf. Human Factors in Computing Systems (CHI)},
  2007.

\bibitem{lavalle2014head}
S.~M. LaValle, A.~Yershova, M.~Katsev, and M.~Antonov.
\newblock Head tracking for the {Oculus} rift.
\newblock In {\em Proc. IEEE Int. Conf. Robotics and Automation (ICRA)}, 2014.

\bibitem{laviola2003double}
J.~J. LaViola.
\newblock Double exponential smoothing: An alternative to {Kalman} filter-based
  predictive tracking.
\newblock In {\em Proc. Eurographics Workshop on Virtual Environments}, 2003.

\bibitem{lee2015outatime}
K.~Lee, D.~Chu, E.~Cuervo, Y.~Degtyarev, S.~Grizan, J.~Kopf, A.~Wolman, and
  J.~Flinn.
\newblock Outatime: Using speculation to enable low-latency continuous
  interaction for mobile cloud gaming.
\newblock In {\em Proc. ACM Int. Conf. Mobile Systems, Applications, \&
  Services (MobiSys)}, 2015.

\bibitem{mark1997post}
W.~R. Mark, L.~McMillan, and G.~Bishop.
\newblock Post-rendering {3D} warping.
\newblock In {\em Proc. Symp. Interactive 3D graphics (I3D)}. ACM, 1997.

\bibitem{hololens}
Microsoft.
\newblock Hololens.
\newblock \url{https://www.microsoft.com/microsoft-hololens}.

\bibitem{miller1968latency}
R.~B. Miller.
\newblock Response time in man-computer conversational transactions.
\newblock In {\em Proc. ACM Fall Joint Computer Conference}, 1968.

\bibitem{monsoon}
Monsoon.
\newblock Monsoon power monitor.
\newblock \\https://www.msoon.com.

\bibitem{mosberger1996making}
D.~Mosberger and L.~L. Peterson.
\newblock Making paths explicit in the {Scout} operating system.
\newblock In {\em Proc. USENIX Conf. Operating Systems Design \& Implementation
  (OSDI)}, 1996.

\bibitem{ng2014blink}
A.~Ng, M.~Annett, P.~Dietz, A.~Gupta, and W.~F. Bischof.
\newblock In the blink of an eye: Investigating latency perception during
  stylus interaction.
\newblock In {\em Proc. ACM Conf. Human Factors in Computing Systems (CHI)},
  2014.

\bibitem{ng2012designing}
A.~Ng, J.~Lepinski, D.~Wigdor, S.~Sanders, and P.~Dietz.
\newblock Designing for low-latency direct-touch input.
\newblock In {\em Proc. Ann. ACM Symp. User Interface Software \& Technology
  (UIST)}, 2012.

\bibitem{nixon2014mobile}
K.~W. Nixon, X.~Chen, H.~Zhou, Y.~Liu, and Y.~Chen.
\newblock Mobile gpu power consumption reduction via dynamic resolution and
  frame rate scaling.
\newblock In {\em 6th Workshop on Power-Aware Computing and Systems (HotPower
  14)}, 2014.

\bibitem{tegrak1}
nVidia.
\newblock Tegra k1 technical reference manual.

\bibitem{g-sync}
NVIDIA.
\newblock G-sync.
\newblock \url{http://www.geforce.com/hardware/technology/g-sync}, 2014.

\bibitem{oculusSensor}
Oculus{ }VR.
\newblock Building a sensor for low latency vr.
\newblock
  \url{https://www3.oculus.com/en-us/blog/building-a-sensor-for-low-latency-vr}.

\bibitem{oculusRift}
Oculus{ }VR.
\newblock Oculus rift.
\newblock \url{https://www3.oculus.com/en-us/rift}.

\bibitem{pasqual2014mouse}
P.~T. Pasqual and J.~O. Wobbrock.
\newblock Mouse pointing endpoint prediction using kinematic template matching.
\newblock In {\em Proc. ACM Conf. Human Factors in Computing Systems (CHI)},
  2014.

\bibitem{fastsync}
T.~Petersen.
\newblock Gpu boost 3 and sli, May 7, 2016.

\bibitem{dsiCommand}
E.~Petillon.
\newblock Demystify {DSI} {I/F}.
\newblock Texas Instruments, 2012.

\bibitem{segal1992opengl}
M.~Segal and K.~Akeley.
\newblock The {OpenGL} graphics system: A specification (version 1.0), 1992.

\bibitem{seow2008designing}
S.~C. Seow.
\newblock {\em Designing and engineering time: The psychology of time
  perception in software}, chapter~3.
\newblock Addison-Wesley Professional, 2008.

\bibitem{instrumentsomap4460}
Texas{ }Instruments.
\newblock Omap4460 multimedia device silicon: Trm.

\bibitem{lesnumeriques_note3}
R.~Thuret.
\newblock Samsung galaxy note 3 : que valent son écran amoled et son appareil
  photo.
\newblock
  \\http://www.lesnumeriques.com/telephone-portable/samsung-galaxy-note-3-que-valent-son-ecran-amoled-son-appareil-photo-a1740.html,
  2013.

\bibitem{iosPrediction}
P.~Tsoi and J.~Xiao.
\newblock Advanced touch input on {iOS}: Increasing responsiveness by reducing
  latency.
\newblock The Apple Worldwide Developers Conference (WWDC), 2015.

\bibitem{1msdelay}
C.~Velazco.
\newblock Microsoft envisions a future with super-fast touchscreens.
\newblock
  \\http://techcrunch.com/2012/03/09/microsoft-demos-super-fast-touchscreen-but-will-they-ever-make-it-to-market,
  2012.

\bibitem{yan2005mascots}
L.~Yan, L.~Zhong, and N.~Jha.
\newblock Towards a responsive, yet power-efficient, operating system: A
  holistic approach.
\newblock In {\em Proc. IEEE Int. Symp. Modeling, Analysis, and Simulation of
  Computer and Telecommunications Systems(MASCOTS)}, 2005.

\bibitem{yan2015optimizing}
Y.~Yan, S.~He, Y.~Liu, and L.~Huang.
\newblock Optimizing power consumption of mobile games.
\newblock In {\em Proceedings of the Workshop on Power-Aware Computing and
  Systems}, pages 21--25. ACM, 2015.

\bibitem{yang2008redline}
T.~Yang, T.~Liu, E.~D. Berger, S.~F. Kaplan, and J.~E.~B. Moss.
\newblock Redline: First class support for interactivity in commodity operating
  systems.
\newblock In {\em Proc. USENIX Conf. Operating Systems Design \& Implementation
  (OSDI)}, 2008.

\end{thebibliography}
